\documentclass[twocolumn,prb,showpacs,preprintnumbers,superscriptaddress,amsmath,amssymb,citeautoscript]{revtex4-1}
\usepackage{graphicx}
\usepackage{dcolumn}
\usepackage{color}
\usepackage{bm}\usepackage[export]{adjustbox}
\usepackage[hidelinks]{hyperref}
\hypersetup{
    colorlinks,
    citecolor=blue,
    filecolor=blue,
    linkcolor=blue,
    urlcolor=blue
}

\newcommand{\im}{\textrm{i}}

\DeclareMathOperator{\sgn}{sgn}\DeclareMathOperator{\e}{e}
\DeclareMathOperator{\C}{\mathcal{C}}
\DeclareMathOperator{\CL}{\mathcal{L}}

\begin{document}

\title{Flat Majorana bands in 2-d lattices with inhomogeneous magnetic fields:\\topology and stability}

\author{N.~Sedlmayr}
\email{nicholas.sedlmayr@cea.fr}
\author{J.M.~Aguiar-Hualde}
\affiliation{Institut de Physique Th\'eorique, CEA/Saclay,
Orme des Merisiers, 91190 Gif-sur-Yvette Cedex, France}
\author{C.~Bena}
\affiliation{Institut de Physique Th\'eorique, CEA/Saclay,
Orme des Merisiers, 91190 Gif-sur-Yvette Cedex, France}
\affiliation{Laboratoire de Physique des Solides, UMR 8502, B\^at. 510, 91405 Orsay Cedex, France}

\date{\today}

\begin{abstract}
In this paper we show that for a range of configurations of inhomogeneous magnetic fields it is possible to create flat bands of Majorana states localized on the edges of 2-d lattices. Majorana bound states have been predicted to exist in both one dimensional and two dimensional systems with Rashba spin-orbit coupling, magnetic fields, and placed in proximity to a superconductor. For the proposed systems we present the topological phase diagrams, and we study the conditions for weak topology which predict the formation of bands of Majorana states. The Majorana bands are demonstrated to be relatively stable with respect to a variety of different perturbations on both square and hexagonal lattices.
\end{abstract}

\pacs{71.70.Ej, 73.20.-r, 73.22.Pr, 74.45.+c}

\maketitle

\section{Introduction}

Majorana fermions are fermionic particles which are their own anti-particle.\cite{Majorana1937} The possibility of creating Majorana bound states in the laboratory in condensed matter systems, as well as their possible application to quantum computing,\cite{Kitaev2001,Nayak2008,Santos2010b} has instigated a large body of work in this area in recent years. The existence of Majorana bound states in spin-orbit coupled wires in proximity with a superconducting substrate has been proposed theoretically\cite{Lutchyn2010,Oreg2010,Alicea2011} and investigated experimentally,\cite{Mourik2012,Deng2012,Das2012,Lee2014} though a definitive confirmation of the existence of the Majorana bound state has not yet been given.  Via the bulk-boundary correspondence the existence of the Majoranas at boundaries can be shown to be related to the topology of the bulk bandstructure.\cite{Ryu2002} There now exists a large variety of theoretical models which possess Majorana states in  1d, quasi 1d, and 2d systems.\cite{Yazdani1997,Nilsson2008,Fu2008,Fu2009,Akhmerov2009,Law2009,Sato2009,Sato2009a,Sato2009b,Sato2010,Sato2010a,Sau2010,Law2011,Stanescu2011,Potter2011,Cook2011,Martin2012,Tanaka2012,Lim2012,Kjaergaard2012,Klinovaja2012a,Klinovaja2013,Klinovaja2013a,Stanescu2013,Nadj-Perge2013,Sau2013,Wong2013,Mizushima2013,Zhang2013,Zhang2013a,Poyhonen2014,Kim2014a,Heimes2014,Wakatsuki2014,Wang2014,Seroussi2014,Deng2014,Wang2014a,Dutreix2014,Dutreix2014a,Nadj-Perge2014,Guigou2014}

In the experimental systems spin-orbit coupling is crucial for the existence of the topologically non-trivial regime.  However, it is also possible to create the necessary physics with a nonuniform magnetic field. In the reference frame of the magnetic field this inhomogeneity is naturally very similar to a spin-orbit interaction. One proposal to realize such inhomogenous field configurations is to deposit magnetic atoms on a superconducting substrate,\cite{Yazdani1997,Choy2011,Nadj-Perge2013,Sau2013,Pientka2013,Poyhonen2014,Kim2014a,Heimes2014} which has also been checked experimentally.\cite{Nadj-Perge2014} The bound states that form at the magnetic impurities, called Shiba states, then form a lattice of non-uniform magnetic moments.\cite{Shiba1968,Sakurai1970} Helical magnetism gives rise to similar phenomena.\cite{Kjaergaard2012,Martin2012,Klinovaja2012a,Klinovaja2013}
Instead of focusing on a particular origin for the inhomogeneous magnetic fields we consider the magnetic inhomogeneity as a free parameter and explore the phase space as a function of possible forms of inhomogeneity. This allows one to locate the phases in which Majorana bound states can form.

We are interested in 2d superconducting systems in the D and BDI class in the usual classification. A gapped superconductor in the topological class D can be in either a topologically trivial or non-trivial phase described in 2d by a $\mathbb{Z}$ invariant\cite{Schnyder2008}.  Without closing the gap it is impossible to change from one phase to the other. It can be demonstrated that a topologically non-trivial system has edge states at the boundaries.\cite{Ryu2002,Qi2006}  In our case these are zero energy Majorana states and always come in pairs. A BDI system on the other hand has, strictly speaking, no topological ground state in 2d. In this case we can map particular configurations to a set of effective 1d BDI wires each of which is characterised by a $\mathbb{Z}$ invariant. The topological ground state is protected only in so far as the effective wires remain independent. For our purposes it is sufficient to consider the parity of the relevant $\mathbb{Z}$ invariant, which is itself a $\mathbb{Z}_2$ invariant which takes the values $\delta=1$ or $\delta=-1$ in the topologically trivial and non-trivial phases respectively.

The existence of multiple edge states can be understood through the concept of `weak' topology. If the system has translational invariance along one direction, with boundaries parallel to this direction, then by performing a 1-d Fourier transform the 2-d system can be decomposed into a set of independent 1-d systems. Each one labelled by the appropriate quantum number. Each of these independent 1-d systems can in principle be either topologically trivial or non-trivial labelled by its own invariant $\delta_n$, and thus, via the bulk-boundary correspondence, either host edge states or not as appropriate. By calculating the bulk topological invariant for each 1-d Hamiltonian we can find the total number of edge states for a 2-d lattice with specific boundaries. The existence of many such edge states, which are not fully protected, we refer to as weak topology. The resulting plots which show the number of Majorana edge states we refer to as weak topological phase diagrams, though strictly speaking it is of course not a phase diagram for the full system.

Multiple Majorana bound states have already been predicted in several systems. Quasi 1-d many-band wires can host a small number of Majorana states at the ends of the wire.\cite{Stanescu2011,Lim2012,Stanescu2013} Ladders and coupled chains also allow one to build up models with multiple Majoranas by combining the edge states existing for wires.\cite{Mizushima2013,Wakatsuki2014,Wang2014,Seroussi2014,Deng2014}
Flat bands of Andreev bound states\cite{Sato2011} have been predicted to occur at the edge of 2-d topological superconductors, and dispersionless bands of Majorana states will occur in the edge of 2-d square lattices provided the spin orbit coupling parallel to the edge can be neglected.\cite{Wang2014} Flat bands have been predicted in several alternative systems such as ($d_{xy}$+$p$)-wave topological superconductors,\cite{Tanaka2010,Yada2011,Queiroz2014} and ($p+\im p$) superconductors.\cite{Wong2013} Flat bands of Majoranas in systems with s-wave coupling have been studied in spin-orbit coupled systems.\cite{Schnyder2011} The set-up we consider here, where spin-orbit coupling is played by the magnetic inhomogeneity of Shiba states has received only limited attention\cite{Nakosai2013} and no detailed examination of the stability of the flat bands or their topology exists. In this paper we address these issues for a range of magnetic inhomogeneity and lattice structures.

The systems we consider have two possible realizations: Magnetic adatoms on a surface, or a 2-d lattice with an externally created inhomogeneous magnetic field. Magnetic adatoms on a superconducting substrate will have a very low hopping parameter and thus a small bandwidth. This allows one to access what would otherwise be very high magnetic fields and chemical potentials, measured in units of the hopping.
In general we show results for a large range of parameters to make the structure of the phase diagrams clear. In the case of real solid state lattices only a small portion would be experimentally accessible. The superconducting gap would also then be orders of magnitude smaller than that used in the figures here. We have chosen the values to aid numerical computation and visualization, but the choice makes no difference to the underlying physics. In comparison to lattices with Rashba coupling, one advantage of the model used here is that Majoranas can be found for much smaller dopings (chemical potentials) and magnetic field strengths.

We find flat bands of Majorana edge states for a wide range of magnetic fields and parameters, in both square and hexagonal lattices. We also check the stability of these bands with respect to perturbations along the edge of a square lattice.

This paper is organized as follows, in Sec.~\ref{sec_model} we introduce the generic model we use for different lattices. In Secs.~\ref{sec_square} and \ref{sec_hexagonal} we look at the existence and stability of flat bands of Majorana states in square and hexagonal lattices respectively. We conclude in Sec.~\ref{conclusions}.

\section{Tight binding model for nanoribbons with rotating magnetic field and induced superconductivity}\label{sec_model}

We start from a Bogoliubov-de-Gennes Hamiltonian on a general two dimensional lattice written in the Nambu basis, $\tilde\Psi_{j}=(\tilde\psi_{j,\uparrow},\tilde\psi_{j,\downarrow},\tilde\psi^{\dag}_{j,\downarrow},-\tilde\psi^{\dag}_{j,\uparrow})^T$ where $\tilde\psi^{\dag}_{j,\sigma}$ creates a particle of spin $\sigma$ at site $j$. We use Pauli matrices $\vec{\bm \sigma}$ for the spin subspace and $\vec{\bm \tau}$ for the particle-hole subspace. The full Hamiltonian is
\begin{equation}\label{hamiltonian}
\tilde H=\tilde H_0+\tilde H_{\rm B}\,,
\end{equation}
where the first term is
\begin{equation}
\tilde H_0=\sum_{j}\tilde\Psi^\dagger_j\left[-\mu{\bm\tau}^z-\Delta{\bm\tau}^x\right]\tilde\Psi_j
-\frac{t}{2}\sum_{\langle i,j\rangle }\tilde\Psi^\dagger_i{\bm \tau}^z\tilde\Psi_{j}\,.
\end{equation}
$\mu$ is the chemical potential, $t$ the hopping strength, and $\Delta$ the induced superconducting pairing. For now we keep the nature of the two-dimensional lattice, which is here encoded in the form of the nearest-neighbor coupling terms $\langle i,j\rangle $, completely general. In this work we will focus on square- and hexagonal-lattice ribbons with two edges, and with periodic boundary conditions imposed along one direction, but the results can be straightforwardly generalized to other types of lattices. We set $t=\hbar=1$ throughout. 

The second term in Eq.~\eqref{hamiltonian} is a Zeeman magnetic field of strength $B$ given by
\begin{equation}\label{hmag}
\tilde H_{\rm B}=B\sum_j\tilde\Psi^\dagger_j\hat{n}_j\cdot\vec{\sigma}_{\sigma\sigma'}\tilde\Psi_j\,,
\end{equation}
which can locally vary its orientation $\hat{n}_j$. Here we take the rotation characteristics and the strength of the magnetic field to be free parameters, whether the physical origin of the field is intrinsic, due to an applied field, the substrate, or to magnetic adatoms. To understand the energy scales in the following we note here that for $B=\Delta=\mu=0$ then the square lattice has a bandwidth of $4t$, while the hexagonal lattice has a bandwidth of $3t$ and a Van-Hove singularity at $0.5t$.

Crucially the Hamiltonians we consider anti-commute with the particle hole operator $\C=\e^{\im\varphi}{\bm \sigma}^y\otimes{\bm \tau}^yK$, where $K$ is complex conjugation and $\varphi$ is an arbitrary phase: $\{\C,H\}=0$. This ensures that any pair of zero energy states in the system can be written as a pair of Majorana states. In addition we normally have the property $\{P,H\}=0$ where $P={\bm \sigma}^z\otimes{\bm \tau}^y$, though its exact form will depend on the magnetic field used. This particular $P$ is valid for a magnetic field confined to the $xz$ spin plane. This additional property means that Majoranas on a single edge of the system do not hybridize and destroy each other.\cite{Wang2014} As we shall see in Sec.~\ref{sec_stab} breaking this symmetry, by using a non-planar magnetic field, has consequences for the stability of the Majorana states.

The first step is to remove the magnetic inhomogeneity by an appropriate gauge transformation\cite{Haldane1988b,Kjaergaard2012,Klinovaja2013} which will introduce an effective spin-orbit coupling into the lattice. If we parameterize the orientation as
\begin{equation}\label{mag}
\hat{n}_j=(\cos\varphi_j\sin\theta_j,\sin\varphi_i\sin\theta_j,\cos\theta_j)\,,
\end{equation}
then a rotation $\tilde\Psi_{j}=\mathbf{T}_{j}\Psi_{j}$ which diagonalizes the magnetic field term such that
\begin{equation}
\mathbf{T}^\dagger_j\,\hat{n}_j\cdot\vec{{\bm \sigma}}\,\mathbf{T}_j={\bm \sigma}^z
\end{equation}
can be written as
\begin{equation}
\mathbf{T}_j=\e^{-\frac{\im\varphi_j}{2}{\bm \sigma}^z} \e^{-\frac{\im\theta_j}{2}{\bm \sigma}^y}\,.
\end{equation}
We begin by assuming that the magnetic field is rotating in a single plane in spin space and along a single orientation in real space, with $\theta_i=\theta_0$ and $\varphi_i=2\pi \vec q\cdot\vec{R}_i+\varphi_0$. As neither $\theta_0$ nor $\varphi_0$ make any difference to the physics under investigation we make the further simplification $\varphi_0=0$ and $\theta_0=\pi/2$, such that the magnetic field is in-plane. The condition $\theta_i=\pi/2$ will be relaxed later. For the most part we focus on the effects of changing the real space direction given by $\vec q$ and we use the phrase `orientation of the magnetic field' to refer to this real space direction.

This gauge transformation results in the Hamiltonian $\tilde H\to H=H_0+ H_{\rm Z}+H_{\rm so}$, where the last term is generated from the kinetic energy: $\tilde H_0\to H_0+H_{\rm so}$. The pairing term is invariant under such a rotation and by construction $\tilde H_{\rm B}\to H_{\rm Z}$, a diagonal homogeneous Zeeman field. More explicitly:
\begin{eqnarray}
H_0&=&\sum_{j}\Psi^\dagger_j\left[-\mu{\bm\tau}^z-\Delta{\bm\tau}^x\right]\Psi_j\\\nonumber&&\qquad
-\frac{t}{2}\sum_{\langle i,j\rangle }\Psi^\dagger_i\cos[\pi\vec q\cdot\vec\delta_{ij}]{\bm \tau}^z\Psi_{j}\,,
\end{eqnarray}
and
\begin{equation}
H_{\rm Z}=B\sum_j\Psi^\dagger_j{\bm\sigma^z}\Psi_j\,.
\end{equation}
$\vec\delta_{ij}$ is the real space vector between nearest neighbors $i$ and $j$. The effective spin-orbit coupling term generated in addition to $H_0$ is
\begin{equation}\label{eff_so}
H_{\rm so}=\frac{t}{2}\sum_{\langle i,j\rangle }\Psi^\dagger_{i}\im{\bm \sigma}^x\sin[\pi\vec q\cdot\vec\delta_{ij}]{\bm\tau}^z\Psi_{j}\,.
\end{equation}
Note that in addition to the spin-orbit term, the gauge transformation also modifies the kinetic energy. This is due to the diagonal and spin symmetric component of scattering from the inhomogeneous Zeeman field.
Although it is possible to write an effective low-energy theory which can capture most features of the Majorana states, the full topological phase diagram information can only be gained by treating the band structure correctly at the appropriate points.

The effective spin-orbit interaction has some different properties from the intrinsic Rashba spin-orbit interaction which is present when inversion symmetry is broken. A Rashba coupling of strength $\alpha$,
\begin{equation}\label{rashba}
H_{\rm R}=\im\alpha\sum_{\langle i,j\rangle }\Psi^\dagger_i\left(\vec{\delta}_{ij}\times{\vec{\bm \sigma}}\right)\cdot\hat{z} {\bm \tau}^z\Psi_{j}\,,
\end{equation}
preserves, up to an appropriate spin rotation, the underlying rotational symmetry of the lattice. However, the effective coupling given in Eq.~\eqref{eff_so}, because of the form of the rotating field, automatically breaks this symmetry.
One consequence of this is that it is possible to construct edges which are parallel to the direction along which the effective spin-orbit acts. This is impossible for a real Rashba interaction. Conversely it is also possible for it to act exactly perpendicular to an edge, again this is not possible for Rashba. As we shall see in Secs.~\ref{sec_square} and \ref{sec_hexagonal} this has consequences for the weak topological phases of the systems.
Due to the different underlying symmetry it is not possible to map the kind of inhomogeneous field we consider to Rashba coupling in 2-d. Rashba coupling breaks $\{P,H\}\neq0$ and thus such a system can not host Majorana flat bands.\cite{Wang2014}

In the following two sections we focus specifically on square and hexagonal lattices as examples of the formation of extended Majorana edge states.

Eqns.~\eqref{hamiltonian} to \eqref{hmag} represent a minimal effective model for the physics we are interested in. Real system are likely to have longer range hopping also present. However the inclusion of weaker longer range hopping makes only a small quantitative difference to the results we show. As we are principally interested in the topological properties of the band, a low energy description suffices. These conclusions only change in the limit where one must consider hopping over a large length scale.\cite{Pientka2013}

\section{The square lattice}\label{sec_square}

A square lattice with Rashba spin-orbit coupling can possess either one or two Majorana states along an edge, but flat bands are forbidden by the symmetries of the problem, see App.~\ref{app_rashba} for a complete discussion.  As we demonstrate in the following sections, for the inhomogeneous magnetic fields under consideration it is possible to have not merely two, but many Majorana states co-existing along the edge in the weak topological phase.

\subsection{Bulk effective topological phase diagram for a lattice with a rotating magnetic field}

For a 2d Hamiltonian in the D symmetry class, the relevant topological invariant $\delta=(-1)^\nu$, where $\nu$ is the Chern number, can be determined by a consideration of the parities of the filled bands at the TRI momenta.\cite{Sato2009a,Sato2009b} For a BDI system properly speaking there is no topological invariant describing the ground state. Nonetheless we find it useful to define an equivalent invariant $\delta$ which describes the existence of band inversion and is related to the topological invariants of effective 1d wires after a suitable Fourier transform. 
The calculation of this bulk property is not significantly different to the parity of the topological invariant for a square lattice with spin-orbit terms.\cite{Fu2007,Sato2009,Sato2009a,Sato2010a,Sato2009b,Sato2010} The Hamiltonian can be written, after a Fourier transform, as $H=\sum_{\vec k}\Psi^\dagger_{\vec k}\mathcal{H}(\vec k)\Psi_{\vec k}$
with
\begin{equation}
\mathcal{H}(\vec k)=\begin{pmatrix}
f(\vec k)+B & \mathbf{\CL}_{\vec k} & -\Delta & 0\\
 \mathbf{\CL}_{\vec k}^*& f(\vec k)-B & 0 & -\Delta\\
 -\Delta  &0 & B-f(\vec k) & \mathbf{\CL}_{\vec k} \\
0&  -\Delta &  \mathbf{\CL}_{\vec k}^*& -f(\vec k)-B
\end{pmatrix}\,.
\end{equation}
 We have 
\begin{equation}
f(\vec k)=-t(\cos[k_x]\cos[\pi q_x]+\cos[k_y]\cos[\pi q_y])-\mu
\end{equation}
with $\vec q$ describing the form of the magnetic field. The effective spin-orbit coupling term is
\begin{equation}
\CL_{\vec k}=-\im t(\sin[k_x]\sin[\pi q_x]+\sin[k_y]\sin[\pi q_y])\,,
\end{equation}
which crucially vanishes at the time reversal invariant (TRI) momenta, $\hat\Gamma_{(1,2,3,4)}=(\{\pi,\pi\},\{\pi,2\pi\},\{2\pi,\pi\},\{2\pi,2\pi\})$ defined by $\hat\Gamma_i=-\hat\Gamma_i+\hat G$ with $\hat G$ a reciprocal lattice vector.

\begin{figure}
\includegraphics*[width=0.96\columnwidth,valign=t]{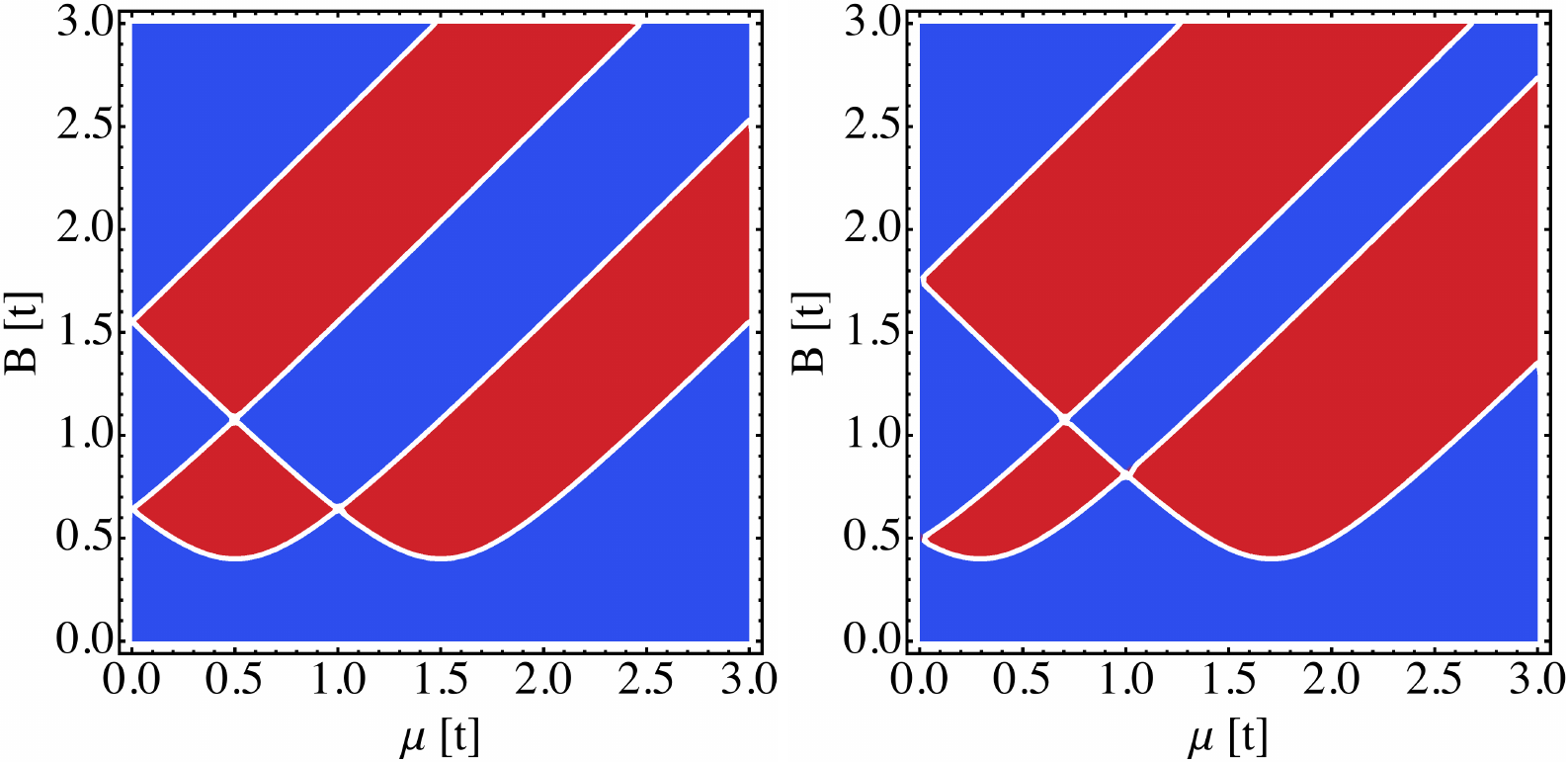}
\caption{(Color online) Bulk phase diagram showing band inversion as a function of $B$ and $\mu$ for a square lattice with $\Delta=0.4$, blue is the $\delta=1$ phase and red the  $\delta=-1$ phase, see main text. Left panel: $\vec q=(1/3,0)$. Right panel:  $\vec q=(3/4,0)$. The phase diagrams are symmetric in $\mu$ and $B$ and hence only positive values are shown.}
\label{top_phase_sq}
\end{figure}

The Hamiltonian at the TRI momenta can be written in block diagonal form as
\begin{equation}
\mathcal{H}(\hat\Gamma_i)=\begin{pmatrix}
\bar{\mathcal{H}}(\hat\Gamma_i)&0\\
0&-\bar{\mathcal{H}}(\hat\Gamma_i)
\end{pmatrix}\,,
\end{equation}
where
\begin{equation}
\bar{\mathcal{H}}(\hat\Gamma_i)=\begin{pmatrix}
B+f(\hat\Gamma_i) & -\Delta\\
 -\Delta&B-f(\hat\Gamma_i)
\end{pmatrix}\,.
\end{equation}
The topological invariant is then
\begin{equation}
\delta=\sgn\prod_{i=1}^4\det\begin{pmatrix}
B+f(\hat\Gamma_i) & -\Delta\\
 -\Delta&B-f(\hat\Gamma_i)
\end{pmatrix}\,.
\end{equation}
When $\delta=-1$ there is band inversion, i.e.~the parity switches between TRI momenta an odd number of times, and for $\delta=1$ there is no band inversion. $\delta=-1$ is therefore equivalent to topologically non-trivial phases and $\delta=1$ to topologically trivial. That there is no true topological invariant can be understood via the bulk boundary theorem as it is always possible to choose edges such that there are no Majorana edge states for either value of $\delta$,

Examples of the bulk `phase' diagram are shown in Fig.~\ref{top_phase_sq}. As can be seen from these figures, a particular phase diagram for $B$ and $\mu$ retains a familiar overall pattern, and shares some features with the bulk topological phase diagram for the Rashba system, see Fig.~\ref{weak_phase_sq_rashba} in App.~\ref{app_rashba}. However the behaviour as a function of the direction and speed of the magnetic field can be rather complex. Here we show results for two exemplary inhomogeneous fields. Clearly for specific $B$, $\mu$, and $\Delta$ it is possible to change the topology simply by changing $\vec q$.

\subsection{The weak topological phase diagram and flat bands}

\begin{figure}
\includegraphics*[width=0.96\columnwidth,valign=t]{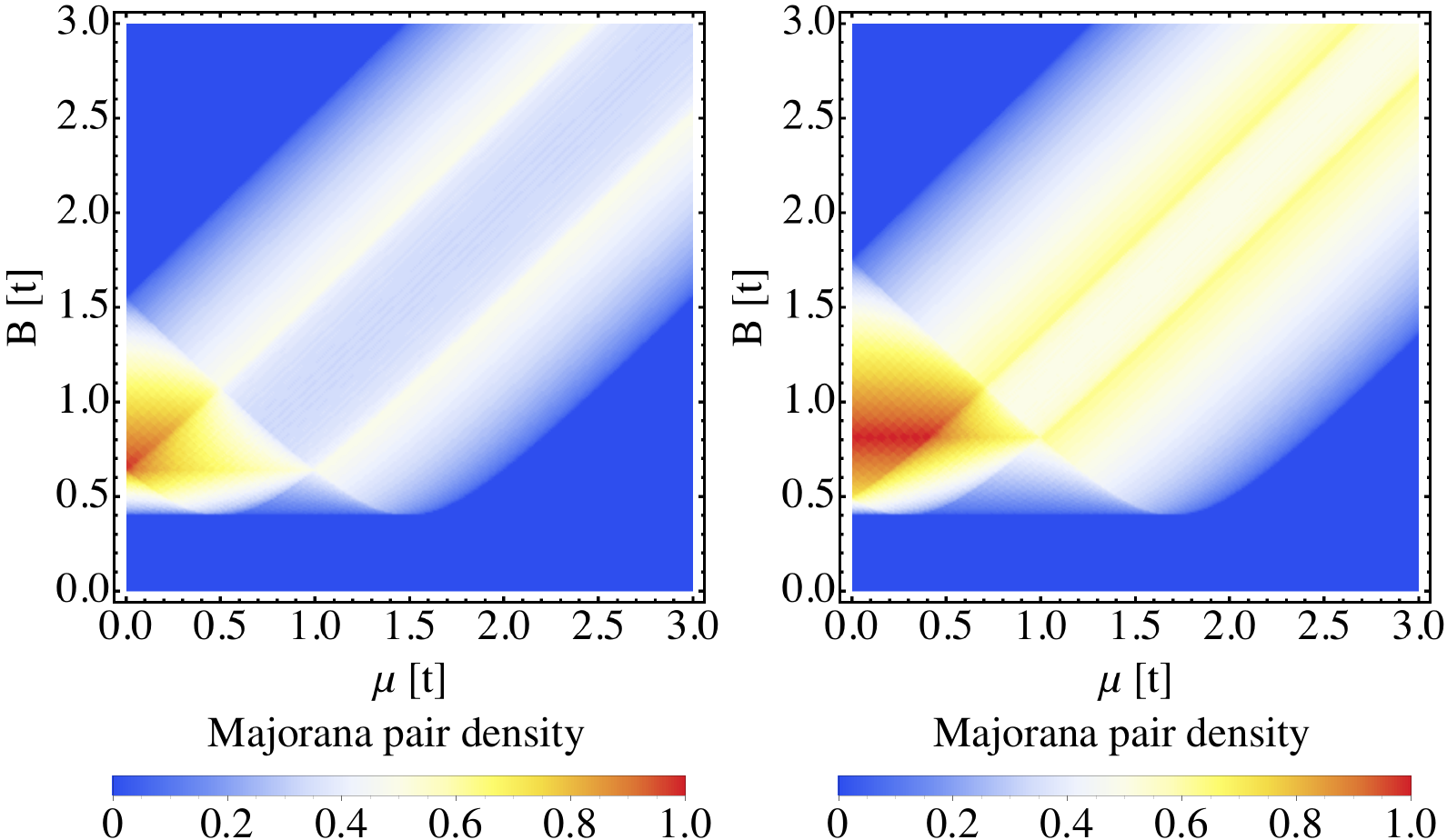}
\caption{(Color online) Weak topological phase diagram as a function of $B$ and $\mu$ for a square lattice with $\Delta=0.4$ showing the density of Majorana pairs. Left panel: $\vec q=(1/3,0)$. Right panel:  $\vec q=(3/4,0)$. Periodic boundary conditions (PBCs) are imposed along $\hat y$ with $N=100$, and with edges running along the same direction.}
\label{weak_phase_sq}
\end{figure}

The form of $\CL_{\vec k}$ that we found for the effective spin-orbit coupling allows us to determine the full weak topological phase diagram in some simple cases. As an example we consider a lattice with straight edges along $x=1$ and $x=N'$.
If we impose periodic boundary conditions along the $y$ direction, perpendicular to the direction of the magnetic field inhomogeneity $\vec q=(q_x,0)$, then we have a set of $N$ effectively independent wires labelled by $k_n=2\pi n/N$ with $n=1,2,\ldots N$. 
Unlike for intrinsic Rashba coupling, see App.~\ref{app_rashba}, we have $\CL_{\pi,k}=\CL_{2\pi,k}=0$  and all independent wires can in principle be topologically non-trivial.
The topological invariant for an independent effective wire is
\begin{eqnarray}
\delta_n&=&\sgn\left[\det\bar{\mathcal{H}}(\pi,k_n)\det\bar{\mathcal{H}}(2\pi,k_n)\right]\\
&=&\sgn\left[B^2-\Delta^2-f^2(\pi,k_n)\right]\left[B^2-\Delta^2-f^2(2\pi,k_n)\right]\,,\nonumber
\end{eqnarray}
with
\begin{equation}
f(k_x,k_n)=-t(\cos[k_x]\cos[\pi q_x]+\cos[k_n])-\mu
\end{equation}
The system can now support many weak topologically non-trivial Majorana bound states along the edges of the square lattice. In Fig.~\ref{weak_phase_sq} we plot the Majorana pair density, defined as $\rho_\gamma\equiv N_\gamma/N$, where $N_\gamma$ is the number of Majorana edge states along a single edge, and $N$ is the total possible number of Majoranas, one pair for each effective wire. Broadly speaking it overlaps the structure familiar from the bulk topological phase diagram. As for the lattice with Rashba coupling there are bulk topologically trivial regions which can support Majorana edge states. The possible number of Majorana edge states here is however much higher.

For a square lattice we can write the Hamiltonian for an effective wire, after the Fourier transform along the longitudinal direction, as
\begin{eqnarray}
\label{SquareHamiltonian}
H_n&=&-\sum_{j=1}^{N'}\Psi^\dagger_{n,j}\bigg\{\left(t\cos\frac{2\pi n}{N}\cos\pi q_y+\mu\right){\bm\tau}^z+\Delta{\bm\tau}^x\nonumber\\
&&-t\sin\frac{2\pi n}{N}\sin\pi q_y\,{\bm\sigma}^y{\bm\tau}^z-B{\bm\sigma}^z\bigg\}\Psi_{n,j}\\&&\nonumber
-\frac{t}{2}\sum_{\langle i,j\rangle }\Psi^\dagger_{n,i}\left(\cos\pi q_x-\im{\bm\sigma}^x\sin\pi q_x\right){\bm\tau}^z\Psi_{n,j}\,.
\end{eqnarray}
The total $2d$ Hamiltonian is $H=\sum_{n=1}^NH_n$.

\begin{figure}
\includegraphics*[width=0.8\columnwidth]{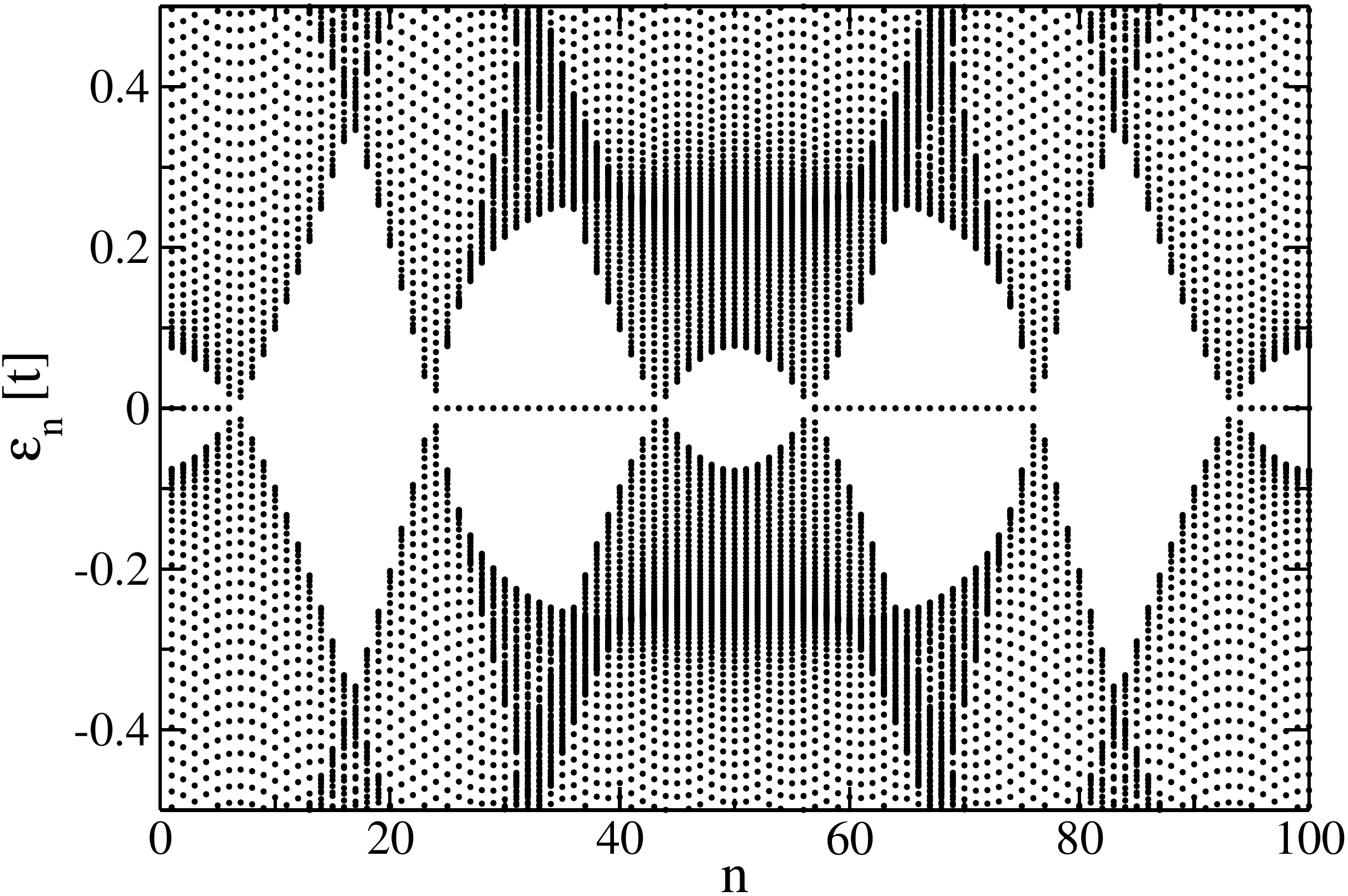}
\caption{A portion of the band structure of a square lattice nanoribbon in the topologically non-trivial regime. Flat bands of Majoranas are clearly visible. $B=1$, $\mu=0.5$, $\vec q=(1/3,0)$, $\Delta=0.4$, and the nanoribbon has a size $N=N'=100$.}
\label{Square_Bandstructure_Rot}
\end{figure}

A typical band structure, in a topologically non-trivial phase, is shown in Fig.~\ref{Square_Bandstructure_Rot}. As for the hexagonal lattice and contrary to the system with spin-orbit coupling, which can support at most two Majorana bound states along an edge, there are flat bands of Majorana edge states. As discussed in the introduction, the weakly protected topology of these Majorana states is connected with the decomposition of the system into a set of independent 1-d systems. In the following section we investigate to what extent these states survive as we perturb the model.

\subsection{Stability of the Majorana bands}\label{sec_stab}

We can now study how the number of Majorana bound states change as we alter the direction of the rotating magnetic field, and as we distort the boundaries. For the square lattice it should be clear that the many weak topological Majoranas depend on there being a component of the magnetic field orientation perpendicular to the edge. If $\vec q$ is parallel to the edge there are of course no Majoranas present.  We will consider three different perturbations to the system. The first is to vary the direction of $\vec q$ to see how quickly the flat bands are destroyed. In this case the momentum resolved picture remains always true and there are two mechanisms by which the flat bands can disappear. Either the bulk gap must close over a range of momenta, or the gap closing points can move, destroying the Majorana states as they go.  This is nothing other than the gap closing and opening for individual momentum $k_n$ resolved systems.

The second type of perturbation we consider is to distort the boundary, causing scattering between the previously uncoupled longitudinal momenta. Although the Majorana edge states do not couple and thus cannot destroy each other, provided they have the appropriate symmetry,\cite{Mizushima2013,Wang2014} scattering can occur between the Majoranas and the finite energy states near the gap closing points. This is one possible mechanism for destroying the flat bands when scattering is allowed between different momenta $k$. As we only distort the edge the bulk states and the bulk topology remains unaffected.
Lastly we will consider the case where $\theta_i$ is not a constant, and vary also the plane in which the magnetic field changes. This breaks the `symmetry' $\{P,H\}=0$ and the Majorana states can now hybridize, destroying each other. In this case we see that the flat bands are much more delicate than for $\theta_i=\theta_0$.

\begin{figure}
\includegraphics*[width=0.96\columnwidth]{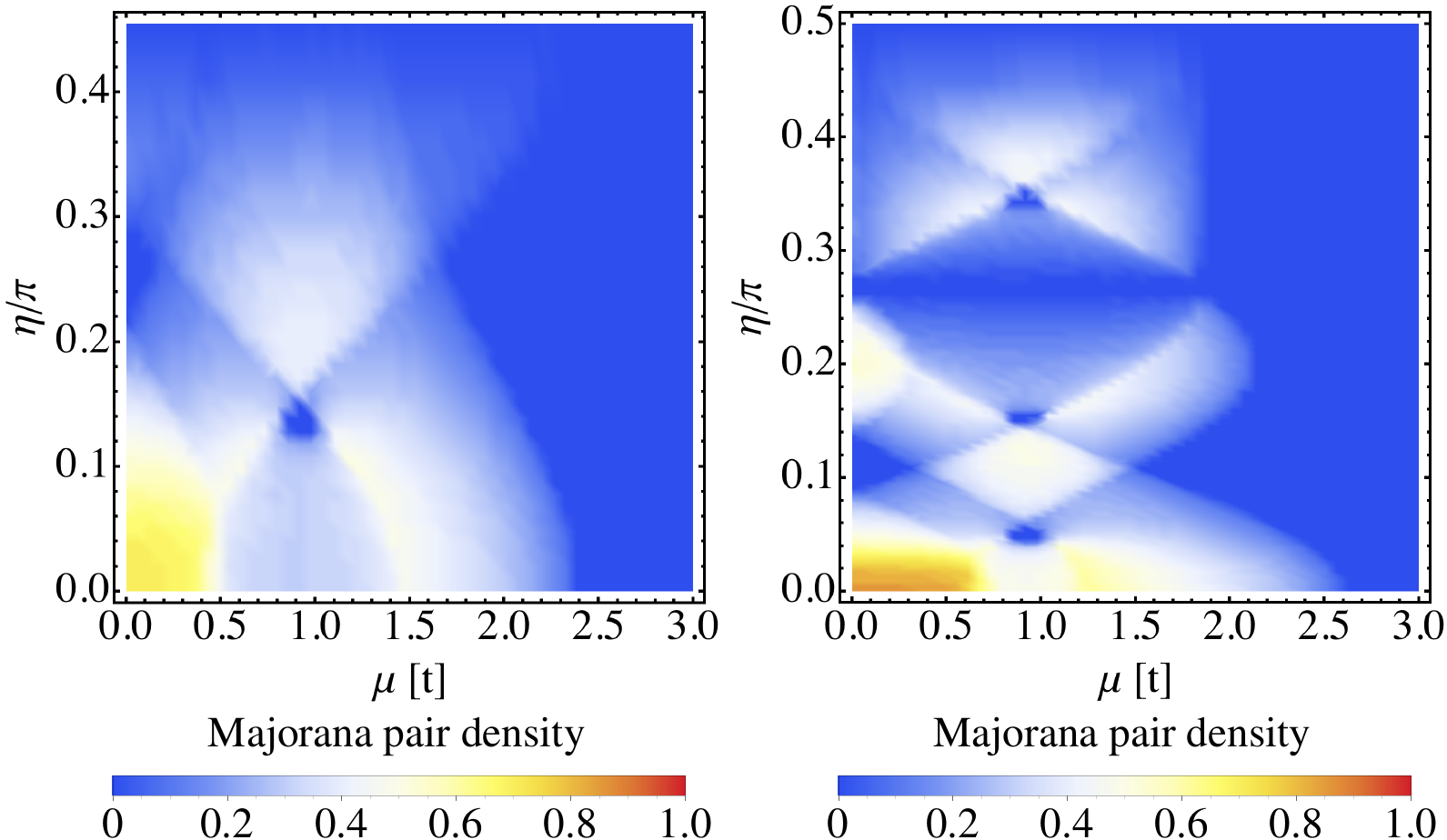}
\caption{(Color online) Numerical results for the density of Majorana pairs, i.e.~the weak topological phase diagram, for a square lattice with open boundary conditions (OBCs) along $\hat x$, periodic boundary conditions (PBCs) along $\hat y$, and with $N=100$, $B=1$, $\Delta=0.4$ and $\vec q=|\vec q|(\cos\eta,\sin\eta)$. On the left $|\vec q|=1/3$ and on the right $|\vec q|=3/4$.}
\label{Weak_Sq_Field}
\end{figure}

In those cases where it is no longer possible to derive the weak topology analytically, we perform numerical simulations. By diagonalizing the tight-binding Hamiltonians for particular lattices we can count the number of zero energy states. `Zero energy' in this case means exponentially small in the transverse system size. It is therefore normally straightforward to insert a cut-off between the nominally zero energy states and the rest of the bands crossing the gap, provided the system is large enough. Fig.~\ref{Weak_Sq_Field} shows results for the Majorana pair density as a function of the chemical potential and orientation of the rotating field direction, $\eta$, where $\vec q=|\vec q|(\cos\eta,\sin\eta)$. As the direction of the rotating magnetic field changes, the number of Majorana states is reduced. For a rotating field orientated parallel to the edge, there can be no Majorana states. However, the Majorana edge states are not very sensitive to small changes in the field orientation and large numbers of edge states can exist for a wide range of rotating fields. For the right hand panel of Fig.~\ref{Weak_Sq_Field}, with $|\vec q|=3/4$, the angles at which the change in the field perpendicular to the edge become commensurate with the lattice show up as additional lines of zero Majorana pair density. For these special values there is effectively no change in $\vec B$ along a line perpendicular to the edge.

By considering the bandstructure as a function of $\eta$ one can see that the slow destruction of Majoranas is related to the shift of the gap-closing points. As this happens different $k_n$ resolved wires undergo the phase transition from topologically non-trivial to trivial, ``destroying'' the Majoranas. The bulk bands are only slowly modified as a function of $\eta$.

\begin{figure}
\includegraphics*[width=0.9\columnwidth]{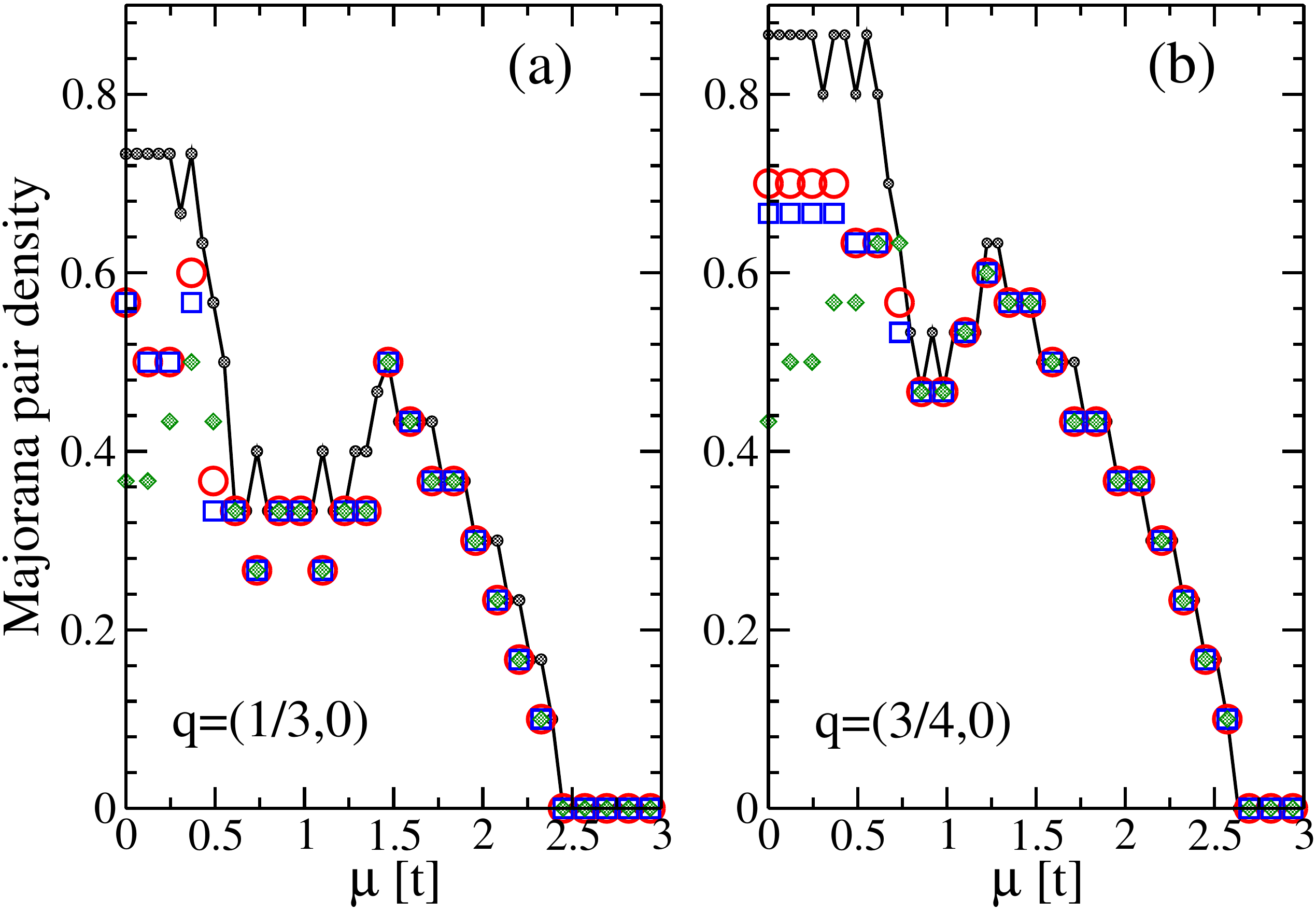}
\caption{(Color online) Numerical results for the Majorana pair density for a square lattice with OBCs $x$ and PBCs along $y$ with $N=30$, $N'=80$, $B=1$, and $\Delta=0.4$. Shown is the Majorana pair density. The solid (black) lines are the analytical results. (Red) circles correspond to a system with a semicircular cut of radius $2.5$ at one edge, the (blue) squares to two semicircular cuts of radius $1.5$ at a distance of $5$ lattice sites, and the (green) diamonds to a ribbon for which every other site on one edge was removed. In both (a) and (b) the bands of Majoranas are stable showing a small reduction in the number of states only for small chemical potentials $\mu\lesssim 0.5$.}
\label{Weak_Sq_cuts}
\end{figure}

Fig.~\ref{Weak_Sq_cuts} shows results for the Majorana pair density as the edge of the system is distorted.
As the edge perturbation does not affect the bulk, these Majoranas can only be destroyed by mixing with the states at the gap closing points in the bandstructure, see Fig.~\ref{Square_Bandstructure_Rot}. Several different types of cut into the edge were tested and all show similar results. For a small chemical potential the disorder induces a drop in the number of Majorana edge states, for intermediate to large $\mu$ the Majoranas seem robust to all types of perturbation we tested.

In general the change in sign of $\delta$, which can be seen in Fig.~\ref{top_phase_sq}, does not have any noticeable effect on the results of Fig.~\ref{Weak_Sq_cuts}, and no change in behaviour is observable as the system crosses between phases. This is expected, as the sign of delta is only related to whether there are an even or odd number of Majorana edge state pairs and this is masked by the large number of weak topological Majorana edge states. Nonetheless, we do observe that the Majorana states are most delicate in the trivial region closest to $\mu=0$.

The weak Majorana states are much less stable if the form of the magnetic field is more complicated, for example if we allow $\theta_j$ to vary, such that the field is no longer confined to a single plane. The gauge field can now introduce scattering between different longitudinal momenta $k_n$ and as $\{P,H\}\neq0$ the Majorana states can hybridize and destroy each other, opening a finite energy gap.\cite{Wang2014} This in turn implies that the Majorana edge states can be more susceptible to changes in the field orientations. If $\theta_j=\pi/2+2\pi \vec \ell_\cdot\vec{R}_j$ then the effective spin-orbit coupling term generated in addition to $H_0$ is now
\begin{equation}\label{eff_so_prec}
H_{\rm so}=-\frac{t}{2}\sum_{\langle i,j\rangle }\Psi^\dagger_{i}{\bm S}^{ij}\Psi_{j}\,,
\end{equation}
where
\begin{eqnarray}\label{eff_so_prec2}
\mathbf{S}^{ij}&=&-\im{\bm \sigma}^z\sin\left[\pi \vec  \ell\cdot\left(\vec R_i+\vec R_j\right)\right]\sin[\pi\vec q\cdot\vec\delta_{ij}]
\nonumber\\&&-\im{\bm \sigma}^x\cos\left[\pi \vec  \ell\cdot\left(\vec R_i+\vec R_j\right)\right]\sin[\pi\vec q\cdot\vec\delta_{ij}]
\\&&+\im{\bm \sigma}^y\sin[\pi\vec  \ell\cdot\vec\delta_{ij}]\cos[\pi\vec q\cdot\vec\delta_{ij}]\,.\nonumber
\end{eqnarray}
The effective spin-orbit terms now depends on the absolute position, not just the relative position.
Similarly the kinetic term is now
\begin{eqnarray}
H_0&=&\sum_{j}\Psi^\dagger_j\left[-\mu{\bm\tau}^z-\Delta{\bm\tau}^x\right]\Psi_j\\\nonumber&&\qquad
-\frac{t}{2}\sum_{\langle i,j\rangle }\Psi^\dagger_i\cos[\pi\vec q\cdot\vec\delta_{ij}]\cos[\pi\vec \ell\cdot\vec\delta_{ij}]{\bm \tau}^z\Psi_{j}\,,
\end{eqnarray}
$H_{\rm Z}$ remains the same.

\begin{figure}
\includegraphics*[width=0.9\columnwidth]{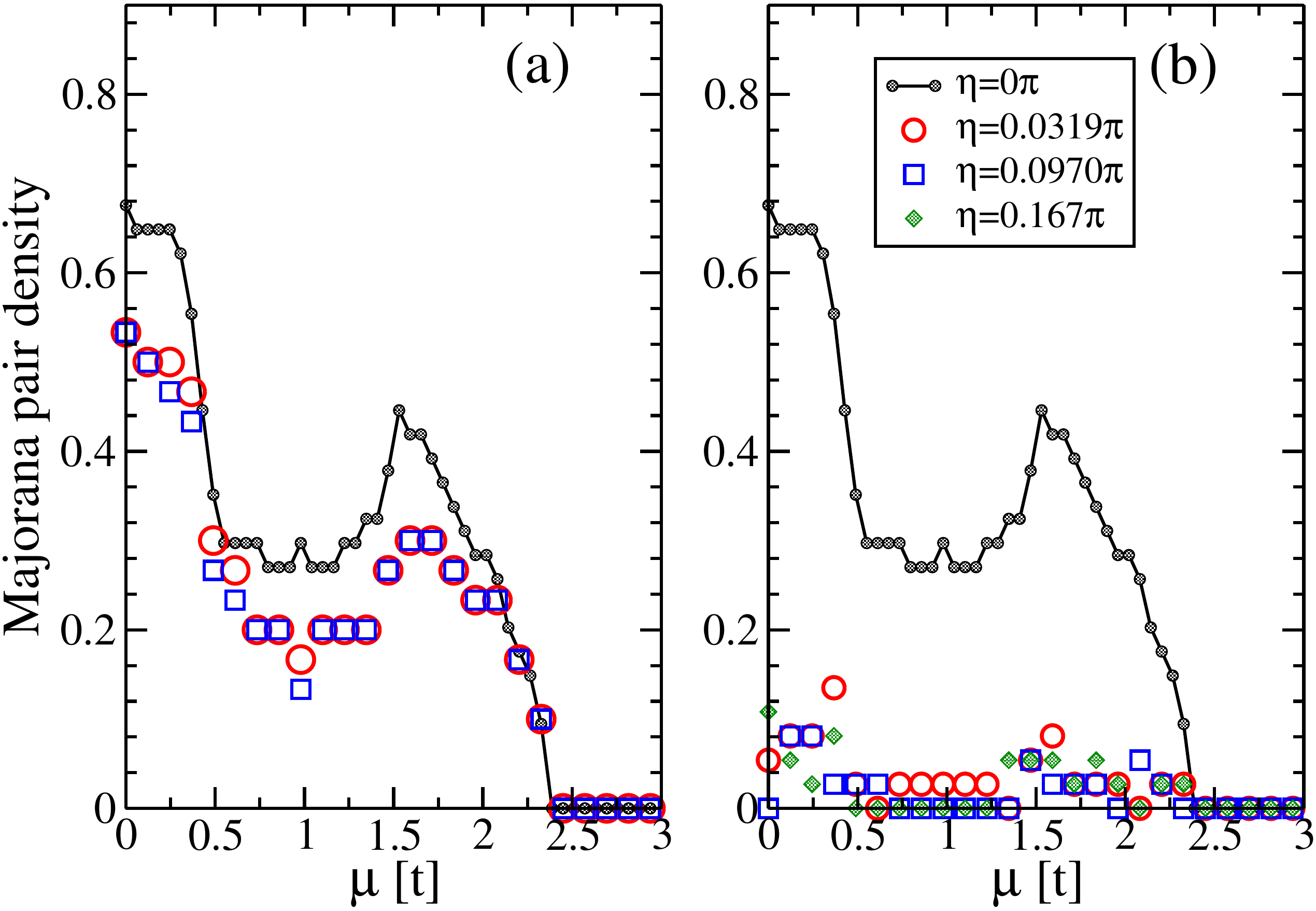}
\caption{(Color online) Numerical results for the density of Majorana pairs for a square lattice as a function of field orientation and chemical potential.  The system has OBCs along $\hat x$ and PBCs along $\hat y$ with $N=30$ and $N'=80$. The other parameters are $B=1$, $\Delta=0.4$ and $\vec q=1/3(\cos\eta,\sin\eta)$ and $\vec \ell=(0.1,0)$.  Panel (a) shows the case for $\eta=0$ and different distortions to the edge, as for Fig.~\ref{Weak_Sq_cuts}(a). Panel (b) shows the density of Majorana pairs as a function of $\eta$, the Majorana flat bands are very quickly destroyed by intraband scattering as $\eta$ is altered.}
\label{Weak_Sq_Field_Prec}
\end{figure}

In Fig.~\ref{Weak_Sq_Field_Prec}(a) we present results for the weak topological phase diagram for $\vec q=|\vec q|(\cos\eta,\sin\eta)$ and $\vec \ell=0.1(\cos\eta,\sin\eta)$. As $\eta$ is varied such that the fields are no longer orientated perpendicular to the edge of the lattice, the Majorana states are very quickly destroyed by intraband scattering. Nonetheless provided we tune to the point where the magnetic field is perpendicular to the edge it is still possible to obtain many Majorana states. Fig.~\ref{Weak_Sq_Field_Prec}(b) shows the effect on the Majorana pair density of cutting into one edge, the states are more delicate than in the case $\theta_j=\pi/2$, compare with Fig.~\ref{Weak_Sq_cuts}(b).

\section{Hexagonal lattice}\label{sec_hexagonal}

The bulk topological phase diagram for isotropic and non-isotropic hexagonal lattices with spin-orbit coupling has been extensively studied before.\cite{Dutreix2014a} In addition there are numerical investigations of the existence and nature of the Majorana edge states in the appropriate phases.\cite{Dutreix2014}  One important difference between the hexagonal and square lattice phase diagrams is the dependence on the Rashba spin-orbit coupling $\alpha$. For a square lattice, neglecting finite size effects, one merely requires $\alpha\neq0$. For a hexagonal lattice the value of $\alpha$ explicitly enters the determination of the topological invariant. Examples of the topological phase diagram for a Rashba hexagonal lattice system are shown in Fig.~\ref{phase_graphene_rashba} in App.~\ref{app_rashba}.
In  contrast to a square lattice, due to the lattice symmetry of the time reversal invariant momenta there is no topologically trivial phase with four Majorana edge states on a hexagonal lattice regardless of the type of edges.

\subsection{Bulk effective topological phase diagram}\label{bulk_hex}

The inhomogenous magnetic fields we use break the symmetry required to calculate the bulk topological order or parity on a hexagonal lattice. The methods used to treat Rashba spin-orbit coupling on the square lattice can be generalized to work on a hexagonal lattice, see App.~\ref{app_rashba} and Ref.~\onlinecite{Dutreix2014a}, but do not help here. Nonetheless one can still find the gap closing points and analyze numerically the parity at the TRI momenta.
After a Fourier transform, see App.~\ref{ft}, the Hamiltonian can be written as $H=\sum_{\vec k}\Psi^\dagger_{\vec k}\mathcal{H}(\vec k)\Psi_{\vec k}$ with
\begin{equation}
\mathcal{H}(\vec k)=\begin{pmatrix}
{\bm f}(\vec k)+B & \CL_{\vec k} & -\Delta & 0\\
-\CL_{\vec k}& {\bm f}(\vec k)-B & 0 & -\Delta\\
 -\Delta  &0 & B-{\bm f}(\vec k) & -\CL^*_{-\vec k} \\
0&  -\Delta &  \CL^*_{-\vec k}& -{\bm f}(\vec k)-B
\end{pmatrix}\,,
\end{equation}
where in the sublattice space
\begin{equation}
{\bm f}(\vec k)=-\frac{t}{2}\sum_{\{\vec\delta\}}\cos[\pi\vec q\cdot\vec\delta]\begin{pmatrix}
0 & \e^{\im \vec k\cdot\vec\delta}\\
 \e^{-\im \vec k\cdot\vec\delta}& 0
\end{pmatrix}\,,
\end{equation}
and
\begin{equation}
\CL_{\vec k}=\frac{\im t}{2}\sum_{\{\vec\delta\}}\sin[\pi\vec q\cdot\vec\delta]\begin{pmatrix}
0 & \e^{\im \vec k\cdot\vec\delta}\\
 \e^{-\im \vec k\cdot\vec\delta}& 0
\end{pmatrix}\,.
\end{equation}
$B$ and $\Delta$ are diagonal in the sublattice space. 
$\vec\delta$ are the nearest neighbor vectors between A and B atoms:
\begin{equation}\label{deltavector}
\{\vec \delta\}=\left\{\left(\frac{\sqrt{3}}{2},-\frac{1}{2}\right),\left(-\frac{\sqrt{3}}{2},-\frac{1}{2}\right),\left(0,1\right)\right\}\,.
\end{equation}
See Fig.~\ref{graphene} for a schematic of the convention used.
Finally we note that
\begin{equation}
\vec k=\left(\frac{2\pi n}{\sqrt{3}N_x},\frac{4\pi m}{3N_y}\right)\,,
\end{equation}
where $n=1,2,\ldots N_x$ and $m=1,2,\ldots N_y$.

\begin{figure}
\includegraphics*[width=0.96\columnwidth]{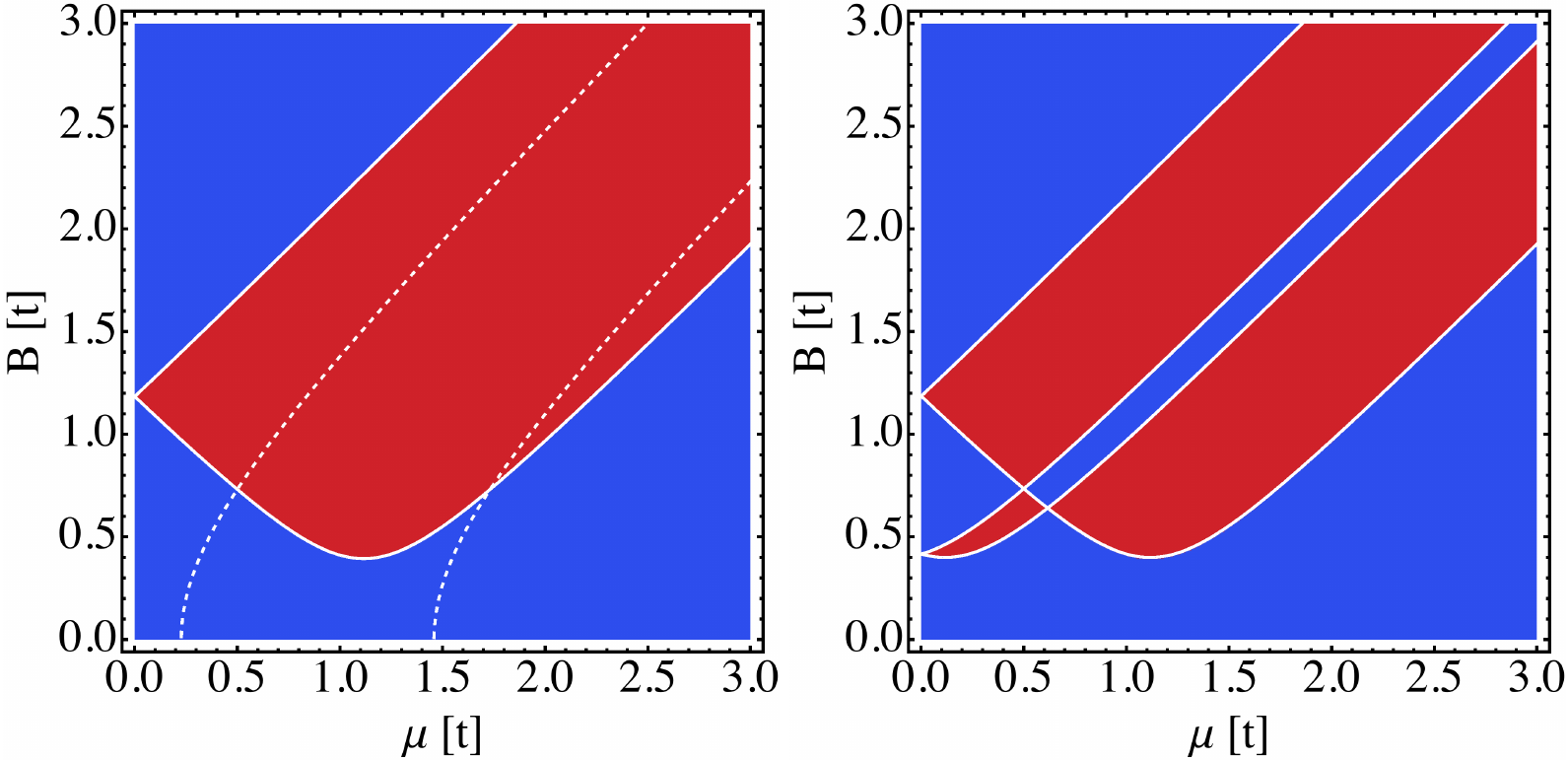}
\caption{(Color online) Bulk phase diagram for a hexagonal lattice with a rotating field along $y$. $\Delta=0.4$ and on the left $\vec q=(0,1/3)$, on the right $\vec q=(1/3,0)$. The solid white lines are the gap closing points separating regions of different parity properties, the dashed white lines are gap closing points between regions of equivalent $\delta$, which has been determined from a numerical analysis of exemplary spectra within the phases: blue is the $\delta=1$ phase and red the  $\delta=-1$ phase.}
\label{phase_graphene_q_y}
\end{figure}

For the hexagonal lattice no simple combination of $\mathbf{\CL}_{\vec k}$ vanishes at all of the TRI momenta. However we can still find the gap closing points which must separate regions of different bulk topology.
The TRI momenta are $\Gamma_i=\{(0,0),(0,2\pi/3),(\pi/\sqrt{3},\pi/3),(\pi/\sqrt{3},-\pi/3)\}$, though it is only necessary to consider $\Gamma_{1,2}$ as $H(\Gamma_{3,4})=H(\Gamma_{2})$.
The bulk topological phase diagram as a function of magnetic field strength and chemical potential, determined from the gap closing points at the TRI momenta, is shown in Fig.~\ref{phase_graphene_q_y}. It has a broadly similar structure to the square lattice phase diagram, Fig.~\ref{top_phase_sq}, but with a different cut into the large $\delta=-1$ phase region, and shows some similarities with the Rashba result, see Fig.~\ref{phase_graphene_rashba}.

\subsection{The weak topological phase diagram and flat bands}

The system also supports many Majorana states in weak topological phases depending on the orientation of the edges and the rotating magnetic field.
For zig-zag edges we can perform the Fourier transform along the $x$ direction, see App.~\ref{ft}, giving $H=\sum_nH_n$ with
\begin{eqnarray}
\label{ZigZagHamiltonian}
H_n&=&\sum_{j=1}^{2N'}\Psi^\dagger_{n,j}\left[-\mu{\bm\tau}^z+B{\bm\sigma}^z-\Delta{\bm\tau}^x\right]\Psi_{n,j}\\&&\nonumber
-t\sum_{j=1}^{2N'-1}\left[\Psi^\dagger_{n,j}\begin{pmatrix}\mathbf{f}^j_{n}&0\\0&-[\mathbf{f}^j_{-n}]^*&\end{pmatrix}\Psi_{n,j+1}+\textrm{H.c.}\right]\,,
\end{eqnarray}
for $n=1,2,\ldots N$. $\mathbf{f}^{j}_{n}$ contains the hopping and spin-orbit terms and is given by
\begin{eqnarray}
\mathbf{f}^{j\in \textrm{even}}_{n}&=&\e^{-\im{\bm \sigma}^x\pi q_y}=\cos[\pi q_y]-\im{\bm \sigma}^x\sin[\pi q_y]\,,\\\nonumber
\mathbf{f}^{j\in \textrm{odd}}_{n}&=&\e^{\frac{\pi n\im}{N}}\e^{\im{\bm \sigma}^x\frac{\pi}{2}(\sqrt{3}q_x+q_y)}+\e^{-\frac{\pi n\im}{N}}\e^{-\im{\bm \sigma}^x\frac{\pi}{2}(\sqrt{3}q_x-q_y)}\,.
\end{eqnarray}

\begin{figure}
\includegraphics*[width=0.8\columnwidth]{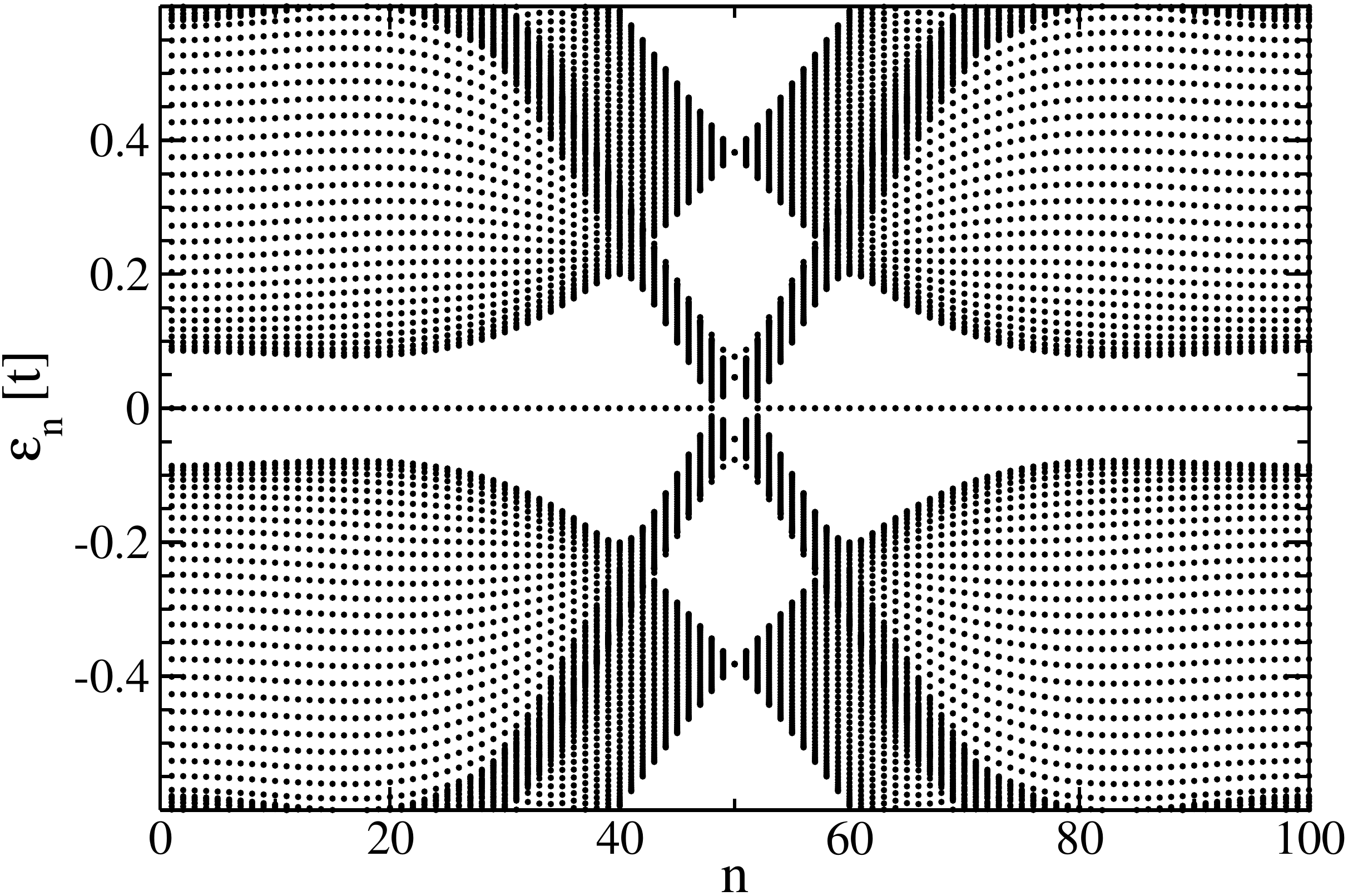}
\caption{A portion of the band structure of a zig-zag edged hexagonal nanoribbon in the topologically non-trivial regime. Flat bands of Majoranas are clearly visible. $B=\mu=1$, $\vec q=(0,1/3)$, $\Delta=0.4$, and the nanoribbon has a size $N=100$ and $N'=50$.}
\label{Graphene_Bandstructure_Rot}
\end{figure}

\begin{figure}
\includegraphics*[width=0.8\columnwidth]{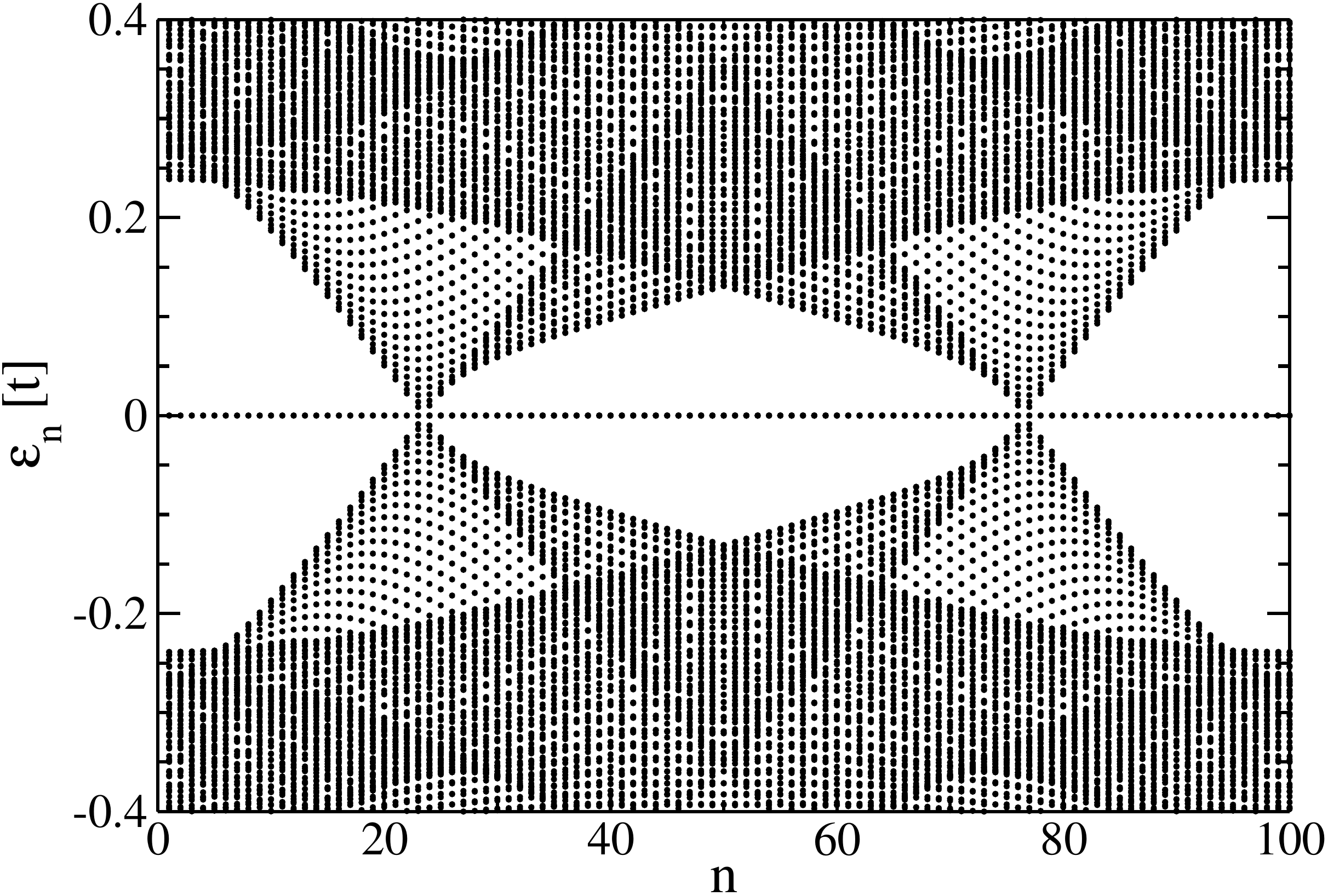}
\caption{A portion of the band structure of an armchair edged hexagonal nanoribbon in the topologically non-trivial regime. Flat bands of Majoranas are clearly visible. $B=1$, $\mu=0.5$, $\vec q=(1/3,0)$, $\Delta=0.4$, and the nanoribbon has a size $N=100$ and $N'=50$.}
\label{Graphene_Bandstructure_Rot_AC}
\end{figure}

A typical band structure, in a topologically non-trivial phase, is shown in Fig.~\ref{Graphene_Bandstructure_Rot} for a zig-zag edge and in Fig.~\ref{Graphene_Bandstructure_Rot_AC} for an armchair edge. Again, as for a square lattice, there are flat bands of Majoranas.

\begin{figure}
\includegraphics*[width=0.7\columnwidth]{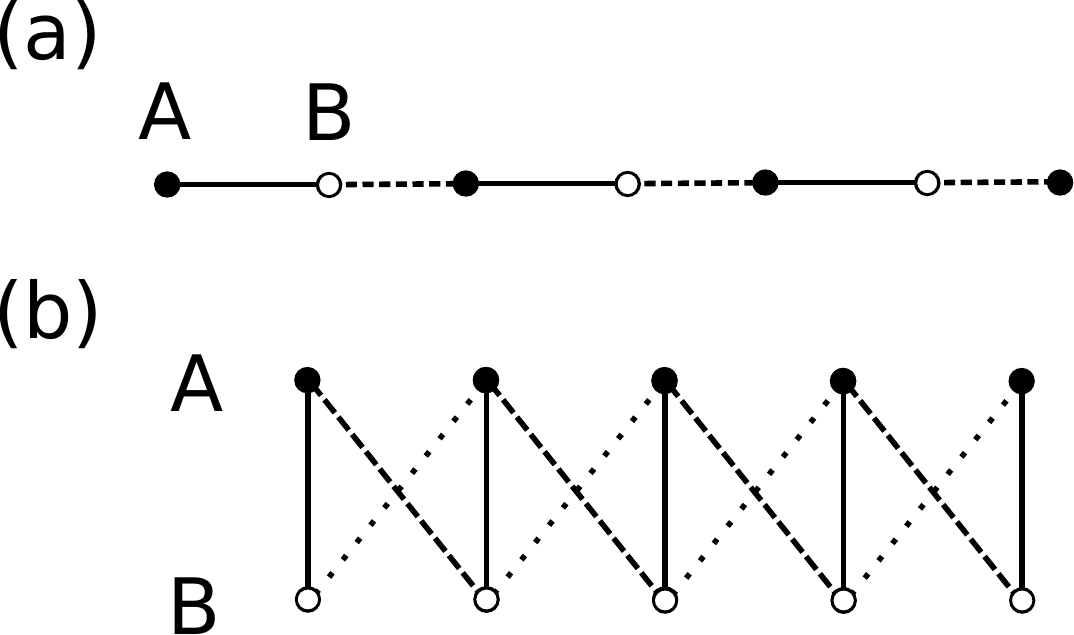}
\caption{Schematics of the lattice structure of the effective wires for system with (a) zig-zag edges, see the Hamiltonian Eq.~\eqref{ZigZagHamiltonian}, and (b) a system with armchair edges, see the Hamiltonian Eq.~\eqref{ACHamiltonian}. Solid, dashed and dotted lines label to different hopping and spin-orbit terms.}
\label{zz_ac_schematics}
\end{figure}

\begin{figure}
\includegraphics*[width=0.96\columnwidth]{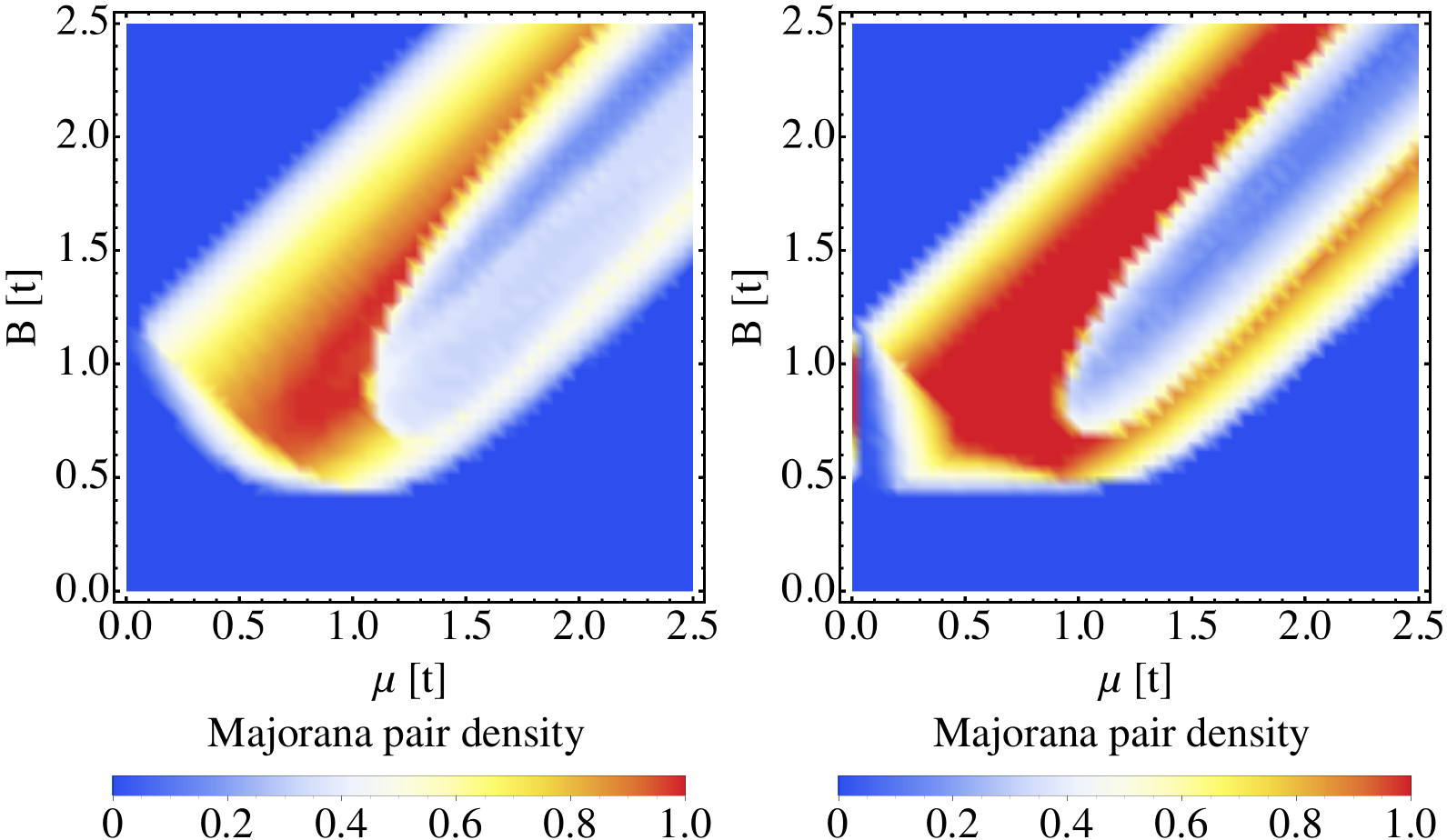}
\caption{(Color online) Numerical results for the density of Majoranas for a hexagonal lattice with $\Delta=0.4$ and $|\vec q|=1/3$, perpendicular to the edge. On the left are results for a nanoribbon with zig-zag edges and with $N'=100$, and $N=100$. Shown is the density of Majorana pairs. On the right is a nanoribbon with armchair edges and $N'=100$ and $N=50$.}
\label{Weak_Hex_BMu}
\end{figure}

For armchair edges we can perform the Fourier transform along the $y$ direction giving $H=\sum_nH_n$ with
\begin{eqnarray}
\label{ACHamiltonian}
H_n&=&\sum_{j=1}^{N'}\sum_{\ell=A,B}\Psi^\dagger_{n,j,\ell}\left[-\mu{\bm\tau}^z+B{\bm\sigma}^z-\Delta{\bm\tau}^x\right]\Psi_{n,j,\ell}\nonumber\\&&
-t\sum_{j=1}^{N'-1}\sum_{\pm}\Psi^\dagger_{n,j,A}\begin{pmatrix}\mathbf{g}^{\pm 1}_{n}&0\\0&-[\mathbf{g}^{\pm 1}_{-n}]^*\end{pmatrix}\Psi_{n,j\pm1,B}\nonumber\\&&\hspace{2cm}+\textrm{H.c.}
\\&&\nonumber
-t\sum_{j=1}^{N'}\Psi^\dagger_{n,j,A}\begin{pmatrix}\mathbf{g}^0_{n}&0\\0&-[\mathbf{g}^0_{-n}]^*\end{pmatrix}\Psi_{n,j,B}+\textrm{H.c.}\,,
\end{eqnarray}
for $n=1,2,\ldots N$ with
\begin{eqnarray}
\mathbf{g}^{0}_{n}&=&\e^{\frac{2\pi n\im}{3N}}\e^{-\im{\bm \sigma}^x\pi q_y}\,,\textrm{ and}\\\nonumber
\mathbf{g}^{\pm1}_{n}&=&\e^{-\frac{\pi n\im}{3N}}\e^{-\im{\bm \sigma}^x\frac{\pi}{2}(\pm \sqrt{3}q_x+q_y)}\,.
\end{eqnarray}
Note that in this case the sublattice is left explicit and labeled by $\ell=A,B$, see App.~\ref{ft}. For such a ladder each independent $H_n$ can support four rather than two Majorana states. In this case we redefine the Majorana pair density as $\rho_\gamma\equiv N_\gamma/2N$. As the system has a $\mathbb{Z}^2$ topological invariant a ladder supporting four Majorana bound states is in a topologically trivial phase; this can have consequences for the stability of the edge sates, as we shall see shortly. Schematics of the effective wire and ladder systems are shown in Fig.~\ref{zz_ac_schematics}. Numerical results for the weak topological phase diagram as a function of $B$ and $\mu$ is shown in Fig.~\ref{Weak_Hex_BMu} for different inhomogeneous fields. The transition to the large region of lower Majorana pair density inside the bulk topological phase is caused by the bulk gap closing over a large range of momenta.

\subsection{Stability of the Majorana bands}\label{sec_stab_hex}

\begin{figure}
\includegraphics*[width=0.96\columnwidth]{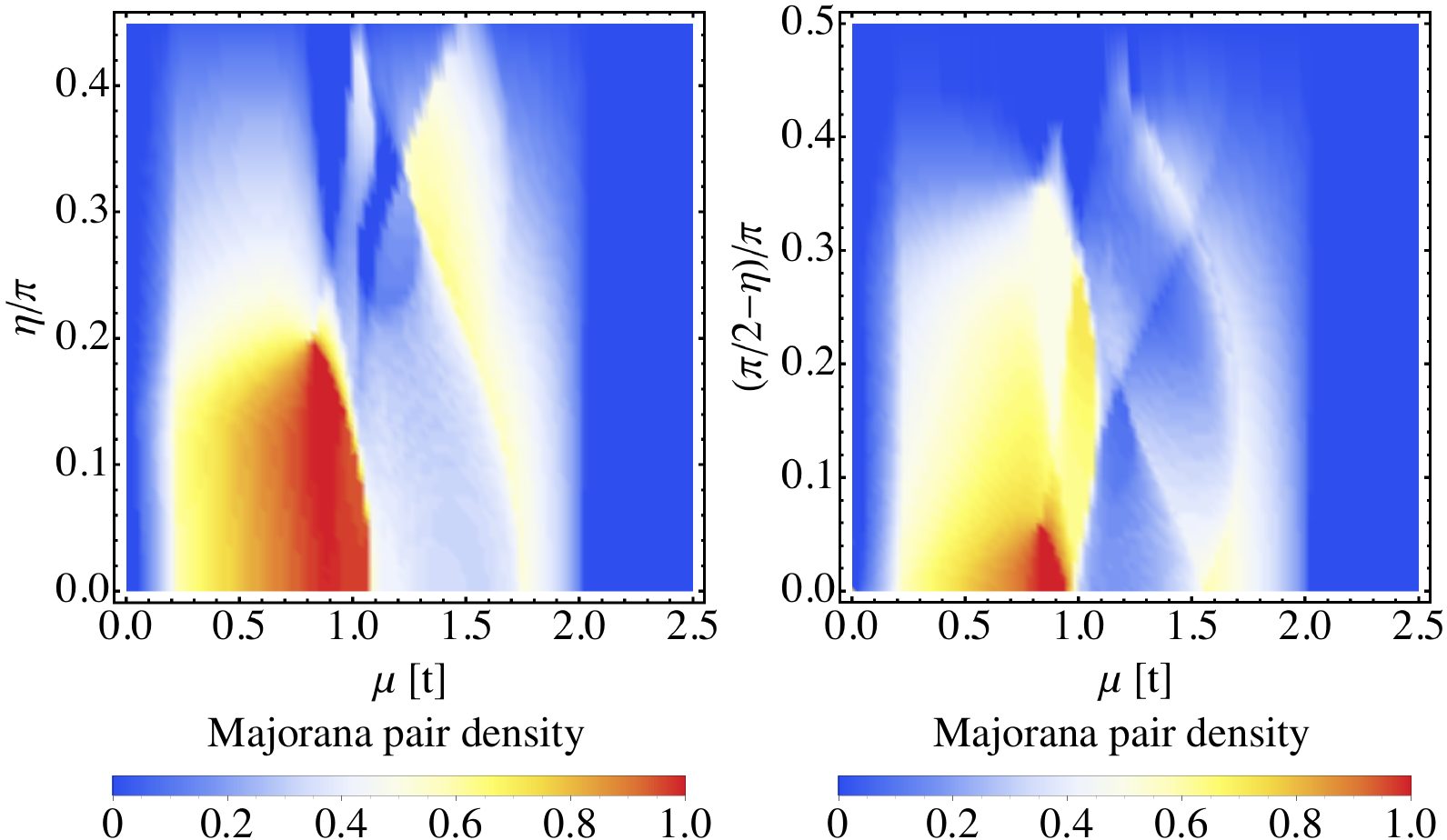}
\caption{(Color online) Numerical results for the density of Majoranas for a hexagonal lattice with $B=1$, $\Delta=0.4$ and $\vec q=1/3(\cos\eta,\sin\eta)$. On the left are results for a nanoribbon with zig-zag edges and with $N'=N=100$. Shown is the density of Majorana pairs. On the right is a nanoribbon with armchair edges and $N'=173$ and $N=58$.}
\label{Weak_Hex_Field}
\end{figure}

Fig.~\ref{Weak_Hex_Field} shows the weak topological phase diagram for a hexagonal lattice with both zig-zag and armchair edges as a function of the magnetic field direction. The zig-zag edges show a stable plateau of edge states which drops suddenly close to $\eta\approx 0.2\pi$. This can be understood by considering the corresponding bandstructures. Changing $\eta$ affects the bulk bands closing the gap; as soon as these bands close a large number of Majorana edge states are destroyed, giving the observed sudden fall-off. For the armchair edge the situation is similar to the square lattice, for which the gap closing points shift as a function of the angle $\eta$, ``destroying'' the Majorana bound states in the process. 

\begin{figure}
\includegraphics*[width=0.96\columnwidth]{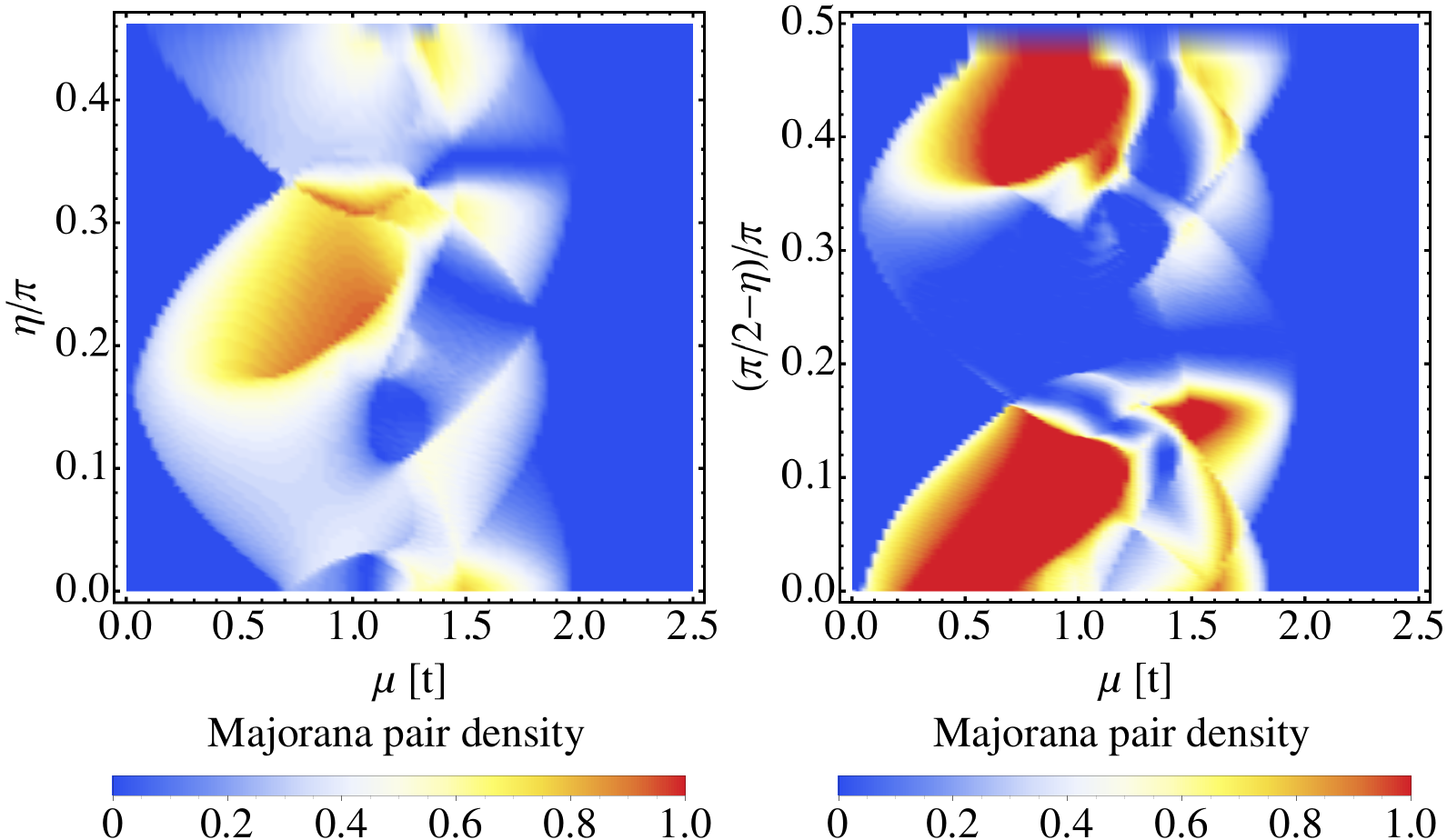}
\caption{(Color online) Numerical results for the density of Majoranas for a hexagonal lattice with $B=1$, $\Delta=0.4$ and $\vec q=3/4(\cos\eta,\sin\eta)$. On the left are results for a nanoribbon with zig-zag edges and with $N'=N=100$. Shown is the density of Majorana pairs. On the right is a nanoribbon with armchair edges and $N'=173$ and $N=58$.}
\label{Weak_Hex_Field2}
\end{figure}

Due to the complicated interplay between the rotating field and the hexagonal lattice, the maximum number of Majorana bound states is not constrained to occur for a magnetic field orientation $\vec q$ perpendicular to the edge. In Fig.~\ref{Weak_Hex_Field2} the weak topological phase diagram is shown for $|q|=3/4$ which has plateaus of large Majorana pair density at various $\eta$.

\begin{figure}
\includegraphics*[width=0.96\columnwidth]{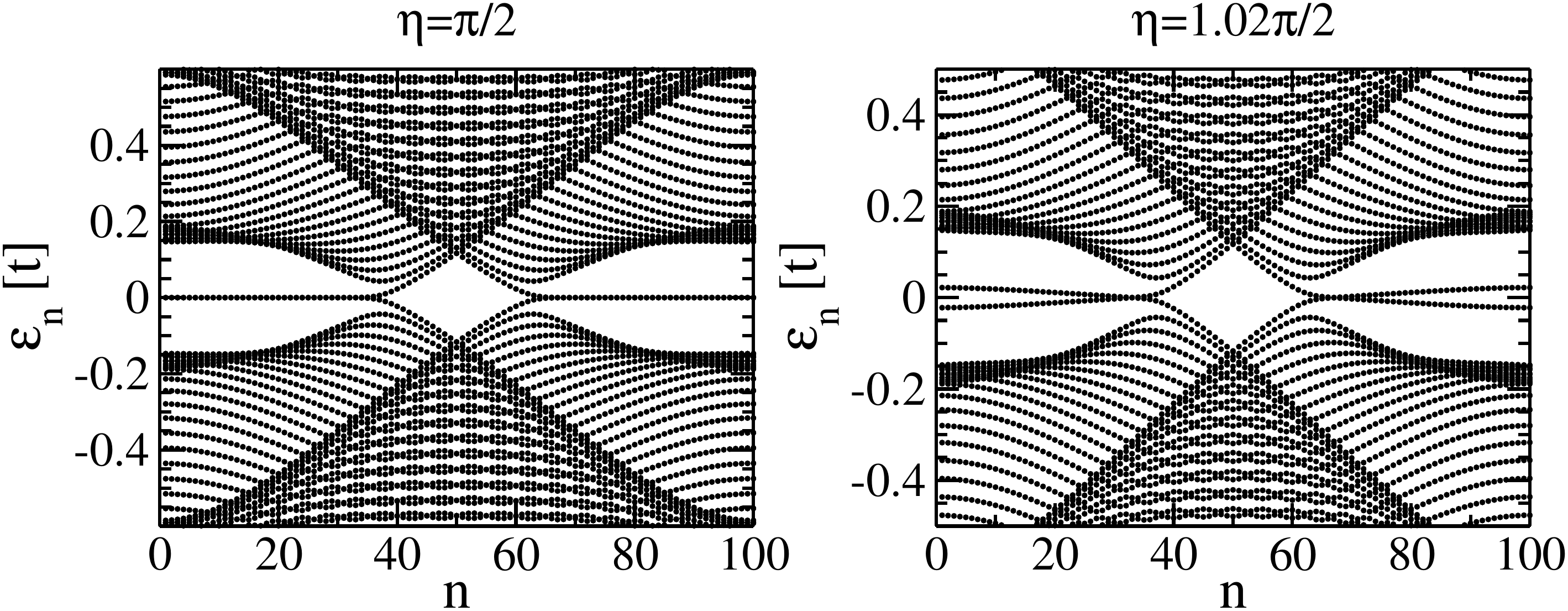}
\caption{A portion of the band structure of an armchair edged hexagonal nanoribbon in the topologically non-trivial regime. Flat bands of Majoranas are clearly visible. $B=1$, $\mu=0$, $\vec q=1/3(\cos\eta,\sin\eta)$ and $\Delta=0.4$; the nanoribbon has a size $N=100$ and $N'=50$. For $\eta=\pi/2$ flat bands of Majorana edge states are visible. These split as $\eta$ is varied, allowed in this case as the Majoranas belong to topologically trivial momentum resolved systems.}
\label{Graphene_Bandstructure_Rot_AC_Mu_0}
\end{figure}

An exception to the previously mentioned ways of destroying the Majorana states can be seen for the feature at $\mu=0$, $\eta=\pi/2$ where a large number of edge states are present. Two pairs of edge states form for each of a set of topologically trivial ladders, this is not visible in Fig.~\ref{Weak_Hex_Field} but can clearly be seen in the bandstructure in Fig.~\ref{Graphene_Bandstructure_Rot_AC_Mu_0}, left hand panel. As these ladders are trivial,  it is not necessary to first close the gap to destroy the zero-energy states, the four zero-energy states are not protected and split spontaneously when the angle $\eta$ is slightly modified from $\pi/2$, becoming non-Majorana localized edge states, see Fig.~\ref{Graphene_Bandstructure_Rot_AC_Mu_0} right hand panel.

\section{Conclusions}\label{conclusions}

We have investigated a generic model of inhomogeneous magnetic fields in two different 2-d lattices: square and hexagonal. In addition to calculating the bulk phase diagrams, which show some interesting distinctions from lattices with spin-orbit coupling, we have explored the weak topological phases. Both lattices exhibit flat bands of Majoranas over a large range of possible parameter values.

We have considered the stability of the flat Majorana bands against various forms of perturbation, which break different symmetries. In those cases where we retain the symmetry $\{P,H\}=0$ then the Majorana states on a single edge do not couple into finite energy states, they are protected. This leaves three possible ways of destroying them. The first is naturally to close the bulk gap. If we do not allow bulk perturbations, such as altering the bulk orientation of the inhomogeneous magnetic field, then there are only two possible mechanisms left. The gap closing points in the 2-d band structure can move, destroying the Majorana zero energy states as they pass, or the Majorana flat band states can scatter from the low energy states near the gap closing points. There is a final possibility belonging to a different class when we have sets of two Majorana states on an edge belonging to a trivial system. These can spontaneously gap themselves out as they have no topological protection, destroying the Majorana states.

If the symmetry protecting the flat bands is broken, then Majorana bands can still exist under special circumstances, but they are very easily destroyed by intraband scattering. 

One perturbation not considered here are electron-electron interactions. Currently there is no universal theory of how topology should be defined in interacting systems, which is a highly non-trivial problem. What happens to the flat bands observed here in the presence of interactions is an interesting open question. One may expect, in analogy to other cases, that the flat bands gain a dispersion, leaving only a single pair of Majorana bound states. In this case the edge states could be amenable to a description as a Luttinger liquid. Alternatively the topological protection may survive for weak interactions, as would be present for the Shiba states under consideration, in which case the Majorana bands could survive. We note that even if the interaction preserves the chiral symmetry protecting the flat bands, as they appear due to a weak topological effect, they are not guaranteed to survive. Answering this question is beyond the scope of the present paper.

The Majorana edge states belonging to the weak topological phases can be seen for lower parameter values than those associated with bulk topological order. This may facilitate their experimental feasibility in comparison with other 2-d lattice models. For lattices composed of Shiba states formed around magnetic adatoms then the effective inter site hopping is much smaller than in a typical lattice and the necessary phases should be feasible. However it should be noted that such lattices may also contain higher order processes not considered in the simple models used here, which could alter the positions of the phase boundaries. An application of these ideas to specific experimental set ups, including all necessary processes, is one possible interesting extension of this work.

\acknowledgements
We would like to thank Pascal Simon, Marine Guigou, and Clement Dutreix for helpful and stimulating discussions.
This work is supported by the ERC Starting Independent Researcher Grant NANOGRAPHENE 256965.

\appendix

\section{Fourier transforms}\label{ft}

In momentum space the Nambu basis vector becomes $\tilde\Psi_{\vec k}=(\tilde\psi_{\vec k,\uparrow},\tilde\psi_{\vec k,\downarrow},\tilde\psi^{\dag}_{-\vec k,\downarrow},-\tilde\psi^{\dag}_{-\vec k,\uparrow})^T$. For the square lattice we use the standard discrete Fourier transform
\begin{eqnarray}
\psi_{j,\sigma}&=&\frac{1}{\sqrt{N}}\sum_{\vec{k}}\e^{\im \vec{k}\cdot\vec{R}_j}\psi_{\vec{k},\sigma}\,,\\\nonumber \vec{k}&=&\left(\frac{2\pi n}{N_x},\frac{2\pi m}{N_y}\right)\,,
\end{eqnarray}
with $n=1,2,\ldots N_x$, $m=1,2,\ldots N_y$, and $N_2=N_xN_y$. $\vec{R}_j=(x_j,y_j)$ is the real space coordinate of lattice site $j$, with a lattice spacing $a=1$.

We also explicitly give here the Fourier transform we use for the hexagonal lattice.
Consider the two dimensional tight binding Hamiltonian
\begin{equation}
H_0=-t\sum_{\langle i,j\rangle ,\sigma}\psi^\dagger_{i\sigma}\psi_{j\sigma}
\end{equation}
defined on a honeycomb lattice. Here $t$ is a hopping matrix element and $\psi^\dagger_{i\sigma}$ creates a particle of spin $\sigma$ on site $i$. If we set periodic boundary conditions (PBCs) in both directions then this can be straightforwardly diagonalized by a Fourier transform.
\begin{figure}
\includegraphics*[width=0.8\columnwidth]{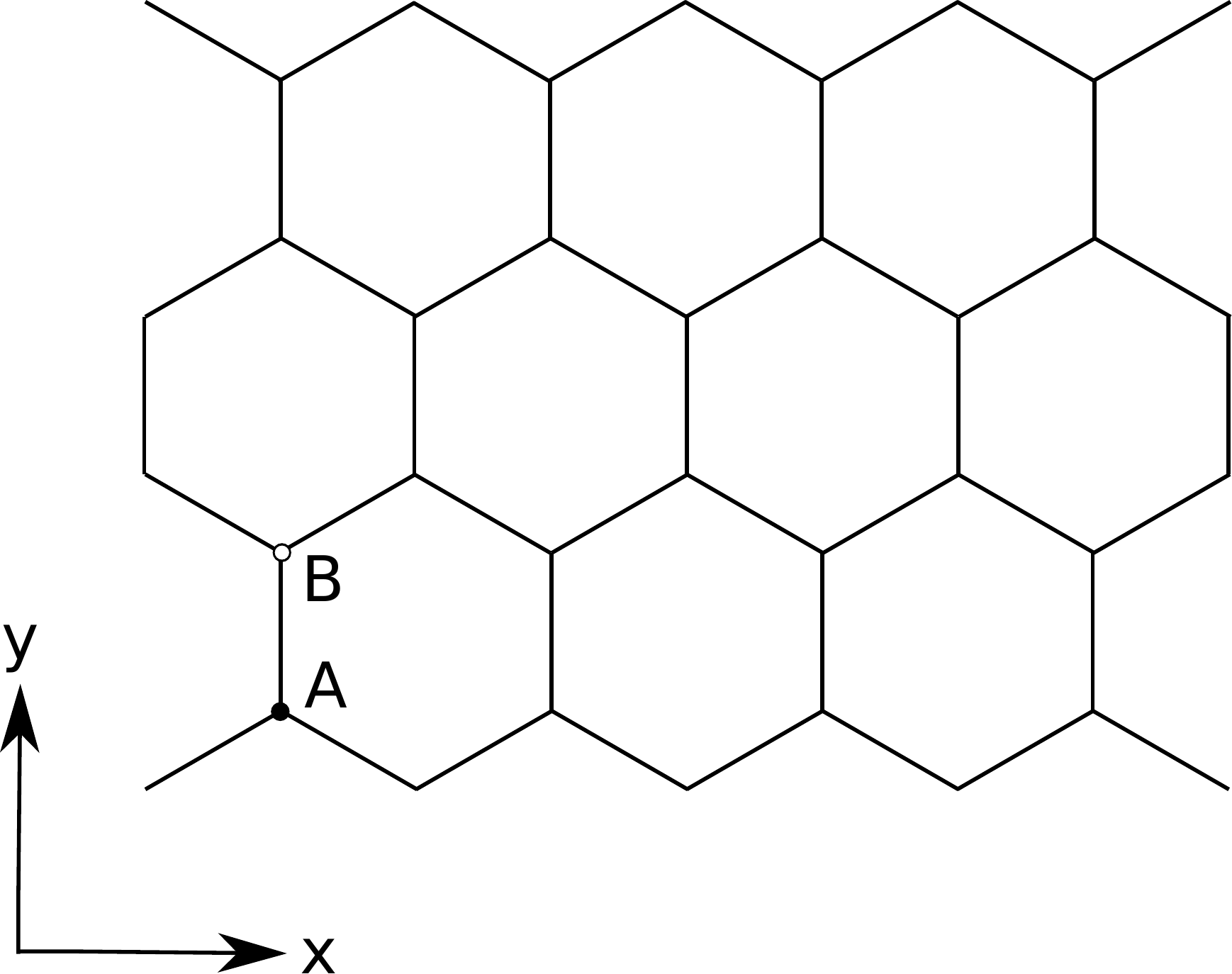}
\caption{A schematic of a small hexagonal nanoribbon with zigzag edges along $\hat x$ and armchair edges along $\hat y$. We stick to this orientation of the nanoribbons throughout this paper.}
\label{graphene}
\end{figure}
Let
\begin{equation}\label{rvector}
\vec{R}(x,y)=\left(\sqrt{3} x+\frac{\sqrt{3}}{2}(1-(-1)^{y}),\frac{3}{2}y\right)
\end{equation}
be the position of the A atom in a unit cell labelled by the integers $x$ and $y$, see Fig.~\ref{graphene}. The nearest neighbors to the B atoms are characterized by the vectors
\begin{equation}\label{deltavector2}
\{\vec \delta\}=\left\{\left(\frac{\sqrt{3}}{2},-\frac{1}{2}\right),\left(-\frac{\sqrt{3}}{2},-\frac{1}{2}\right),\left(0,1\right)\right\}\,.
\end{equation}
The Hamiltonian can be rewritten as
\begin{equation}
H_0=-t\sum_{x,y,\{\vec \delta\},\sigma}c^\dagger_{\sigma}(\vec{R}(x,y))c_{\sigma}(\vec{R}(x,y)+\vec\delta)+\textrm{H.c.}\,.
\end{equation}

For standard periodic boundary conditions the Fourier transform is
\begin{eqnarray}
\psi_{\sigma}(\vec{R}(x,y))&=&\frac{1}{\sqrt{N_2}}\sum_{\vec{k}}\e^{\im \vec{k}\cdot\vec{R}(x,y)}\psi_{\vec{k},\sigma}\,,\\\nonumber \vec{k}&=&\left(\frac{2\pi n}{N_x\sqrt{3}},\frac{4\pi m}{3N_y}\right)\,,
\end{eqnarray}
with $n=1,2,\ldots N_x$, $m=1,2,\ldots N_y$, and $N_2=N_xN_y$ is the number of unit cells.

The Fourier transforms for open boundaries along one direction and periodic along the orthogonal direction are slightly different due to the different periodicity conditions. When considering a Fourier transform along $\hat y$ for a system with open boundary conditions along $\hat x$, i.e.~an armchair edge, then the appropriate momenta are $k=2\pi m/(3N)$ with $N=N_y/2$ the number of repeated armchairs along the edge. Here
$N'=2N_x$ is the number of sites in each effective wire. In this case we leave the sublattice explicit. For a Fourier transform along $\hat x$, applied to a system with open boundary conditions along $\hat y$, the zig-zag edged case, then the appropriate momenta are $k=2\pi n/(\sqrt{3}N)$ with $N=N_x$ the number of repeated zig-zags along the edge, and $N'=N_y$ the number of sites in an effective wire.
It is then convenient to relabel the sublattice index such that the B sites are given by $\Psi_{n,\sigma,2j+1}$ and the A sites are given by $\Psi_{n,\sigma,2j}$ for $j=1,2,\ldots N'/2$.

\section{Topological phase diagram for intrinsic spin-orbit couplings}\label{app_rashba}

First we will recap what is known for the standard square lattice case.\cite{Kitaev2001,Sato2009,Sato2009a,Sato2010a,Sato2010} We will start with the square lattice with Rashba spin-orbit interactions:
\begin{equation}
H=H_0+H_{\rm Z}+H_{\rm R}\,,
\end{equation}
with $H_{\rm R}$ given by Eq.~\eqref{rashba}.
In $H_{\rm Z}$ we set $\vec q=(0,0)$, i.e.~a homogeneous Zeeman field orientated along the $z$ direction.
\begin{figure}
\includegraphics*[width=0.48\columnwidth]{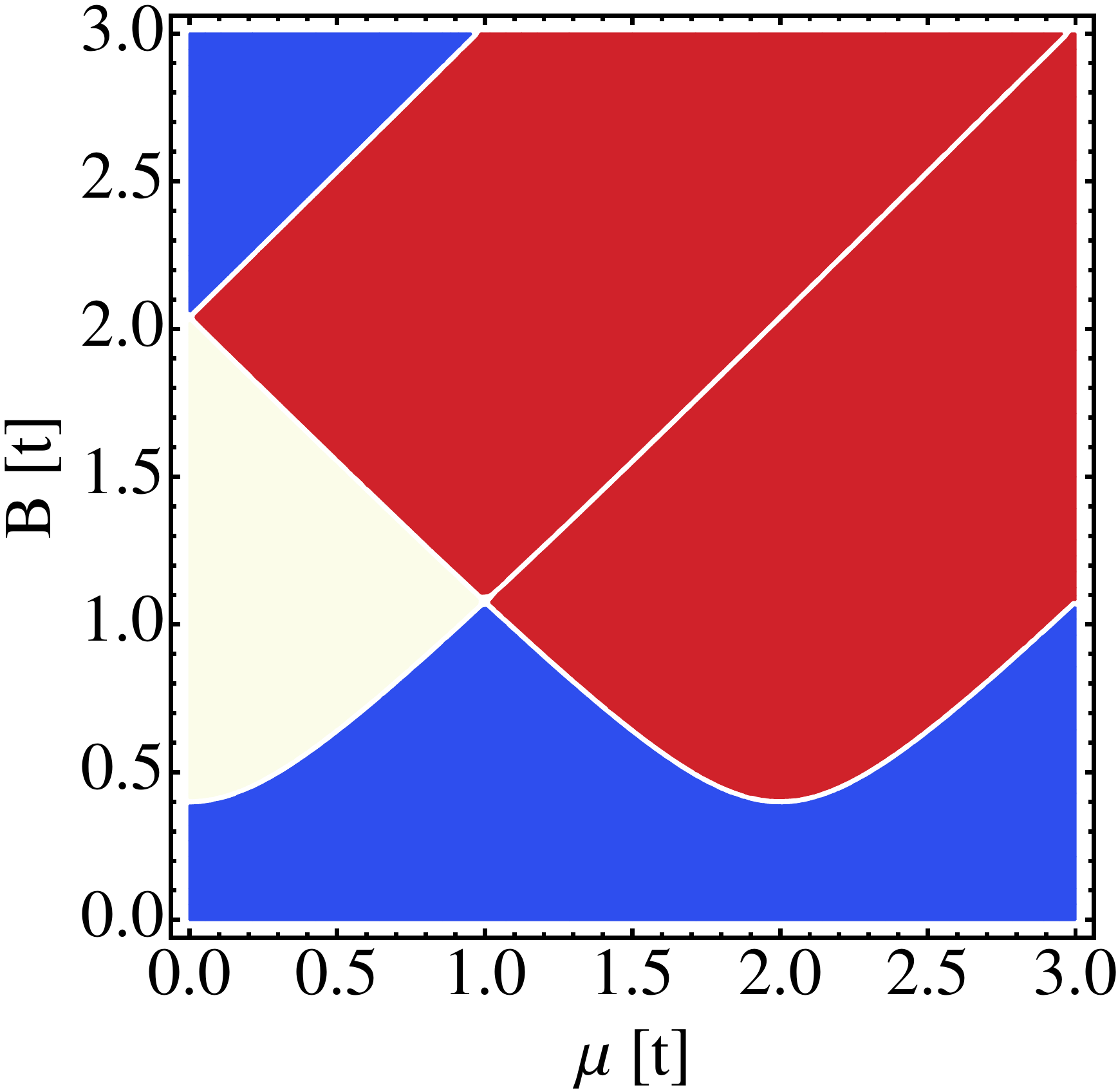}
\caption{(Color online) Bulk topological phase diagram for a square lattice with Rashba coupling and PBCs along $y$, $\Delta=0.4$. Blue is the topologically trivial phase with no Majorana edge states, red is the bulk topologically non-trivial regime with a pair of Majorana edge states and white is the topologically trivial phase with two pairs of Majorana edge states.}
\label{weak_phase_sq_rashba}
\end{figure}

After a Fourier transform with periodic boundary conditions (PBCs) imposed in both directions, see App.~\ref{ft}, the system can be written as $H=\sum_{\vec k}\Psi^\dagger_{\vec k}\mathcal{H}(\vec k)\Psi_{\vec k}$ with
\begin{equation}
\mathcal{H}(\vec k)=\begin{pmatrix}
f(\vec k)+B & \CL_{\vec k} & -\Delta & 0\\
 \CL_{\vec k}^*& f(\vec k)-B & 0 & -\Delta\\
 -\Delta  &0 & B-f(\vec k) & \CL_{-\vec k}^* \\
0&  -\Delta &  \CL_{-\vec k}& -f(\vec k)-B
\end{pmatrix}\,,
\end{equation}
where
\begin{equation}
f(\vec k)=-t(\cos[k_x]+\cos[k_y])-\mu\,,
\end{equation}
and
\begin{equation}
\CL_{\vec k}=-\im\alpha\sin[k_x]-\alpha\sin[k_y]
\end{equation}
is the spin orbit term which vanishes at the four time-reversal-invariant (TRI) points $\hat\Gamma_{(1,2,3,4)}=(\{\pi,\pi\},\{\pi,2\pi\},\{2\pi,\pi\},\{2\pi,2\pi\})$.

The Hamiltonian at the TRI momenta can be written in block diagonal form as
\begin{equation}
\mathcal{H}(\hat\Gamma_i)=\begin{pmatrix}
\bar{\mathcal{H}}(\hat\Gamma_i)&0\\
0&-\bar{\mathcal{H}}(\hat\Gamma_i)
\end{pmatrix}\,,
\end{equation}
where in this case
\begin{equation}
\bar{\mathcal{H}}(\hat\Gamma_i)=\begin{pmatrix}
B+f(\hat\Gamma_i) & -\Delta\\
 -\Delta&B-f(\hat\Gamma_i)
\end{pmatrix}\,.
\end{equation}
The topological invariant is then, provided $\alpha\neq0$,
\begin{eqnarray}\label{rashba_ti}
\delta&=&\prod_i\sgn\left[\det\bar{\mathcal{H}}(\hat\Gamma_i)\right]=\sgn\left[\det\bar{\mathcal{H}}(\hat\Gamma_1)\det\bar{\mathcal{H}}(\hat\Gamma_4)\right]\nonumber\\
&=&\sgn\left\{\left[B^2-\Delta^2-f^2(\hat\Gamma_1)\right]\left[B^2-\Delta^2-f^2(\hat\Gamma_4)\right]\right\}\,.\nonumber\\&&
\end{eqnarray}
When $\delta=-1$ the system is topologically non-trivial and hosts Majorana states on the edges, for $\delta=1$ it is topologically trivial.

Note however that if we impose periodic boundary conditions along for example $y$, then we have a set of effectively independent wires labelled by $k=2\pi n/N$ with $n=1,2,\ldots N$. There are one or two of these wires wires for which $\CL_{\vec k}$ still vanishes at the relevant TRI momenta, which for a simple wire are $k_x=\pi,2\pi$. These are
\begin{eqnarray}
\delta_\pi=\sgn\left[\det\bar{\mathcal{H}}(\hat\Gamma_1)\det\bar{\mathcal{H}}(\hat\Gamma_3)\right]\hspace{3.5cm}\\
=\sgn\left\{\left[B^2-\Delta^2-f^2(\hat\Gamma_1)\right]\left[B^2-\Delta^2-f^2(\hat\Gamma_3)\right]\right\}\,.\nonumber
\end{eqnarray}
for $k=\pi$, which is only present for even $N_y$, or for $k=2\pi$:
\begin{eqnarray}
\delta_{2\pi}=\sgn\left[\det\bar{\mathcal{H}}(\hat\Gamma_2)\det\bar{\mathcal{H}}(\hat\Gamma_4)\right]\hspace{3.5cm}\\
=\sgn\left\{\left[B^2-\Delta^2-f^2(\hat\Gamma_2)\right]\left[B^2-\Delta^2-f^2(\hat\Gamma_4)\right]\right\}\,.\nonumber
\end{eqnarray}
As such it is possible for the system to host four Majorana states in the topologically trivial regime (as predicted by a decomposition of the bulk invariant), $\delta=\delta_\pi\delta_{2\pi}=1$ with $\delta_\pi=\delta_{2\pi}=-1$. All other effective wires are trivial and any Majorana states must occur at the TRI momenta.
Fig.~\ref{weak_phase_sq_rashba} shows the weak topological phase diagram for non-zero Rashba coupling.
Note that the weak topological phase diagram is not a real phase diagram, but rather a diagram which show the number of Majorana edge states, as predicted by the phase diagrams of the underlying 1-d systems.

\begin{figure}
\includegraphics*[width=0.96\columnwidth]{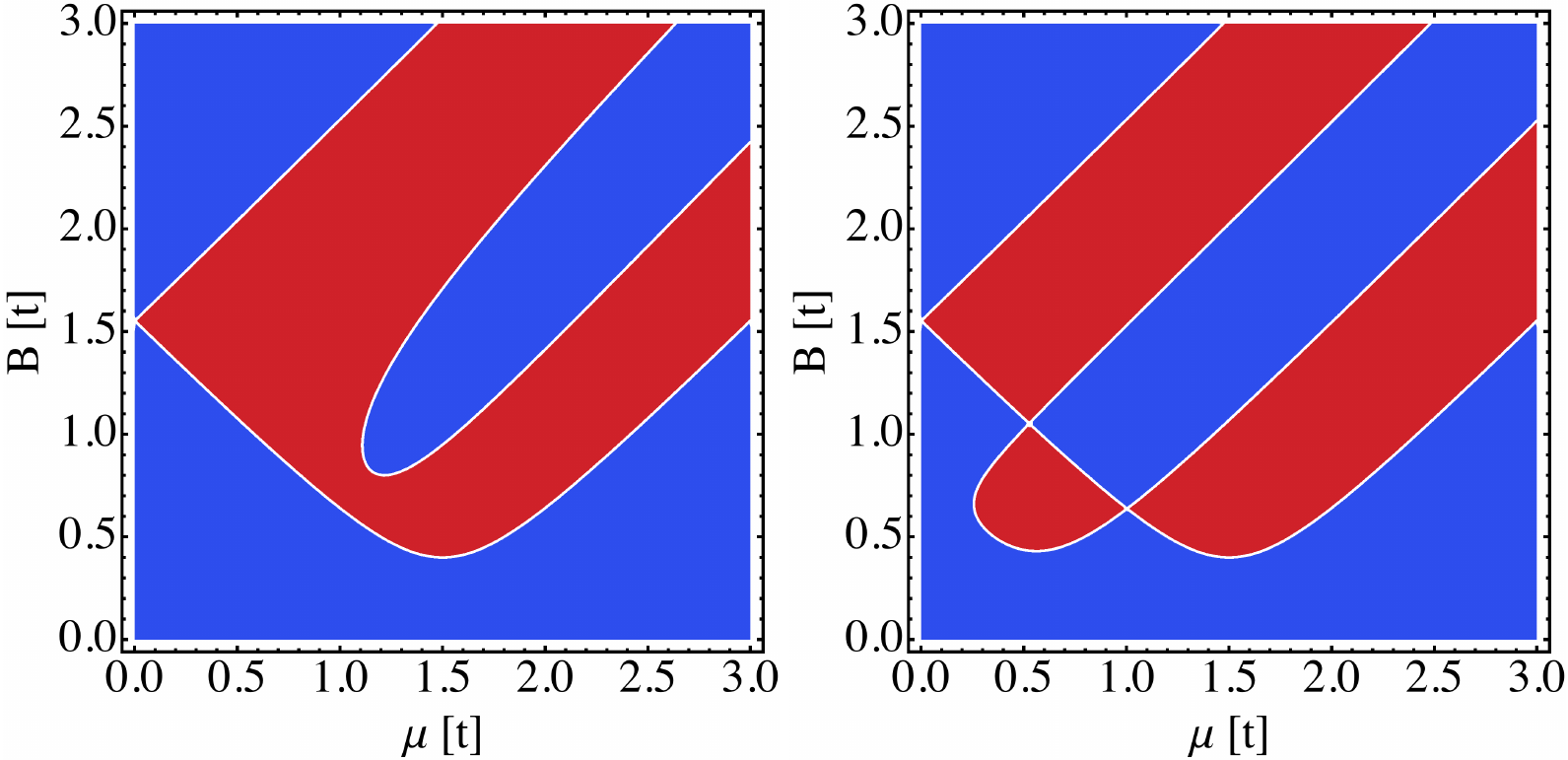}
\caption{(Color online) Bulk topological phase diagram for a hexagonal lattice with Rahsba spin orbit interaction. On the left hand side $\Delta=0.4$ and $\alpha=\sqrt{3}/4$, comparable to a rotating field with $|\vec q|=1/3$. On the right we take $\Delta=0.4$ and $\alpha=0.1$, showing the Majorana states to be found near the Van-Hove singularity and the bottom of the band.}
\label{phase_graphene_rashba}
\end{figure}

In Fig.~\ref{phase_graphene_rashba} we show the bulk topological phase diagram for a hexagonal lattice with Rashba coupling, dealt with in detail in Ref.~\onlinecite{Dutreix2014a}. This is calculated as for the square lattice but with an additional rotation for the sublattice which imposes the appropriate symmetry of the spin-orbit coupling terms. The Hamiltonian at three of the TRI momenta is identical, as for the inhomogeneous magnetic field case discussed in Sec.~\ref{bulk_hex}. As Majorana states can only exist here at the TRI momenta, there is no possibility to have more than a single pair of Majorana state solutions.

\end{document}